\documentclass[twocolumn,amsmath,nofootinbib]{revtex4} 

\usepackage{url}
\usepackage{amssymb}
\usepackage{amsfonts}
\usepackage{amsbsy}
\usepackage{graphicx}
\usepackage{hyperref}
\usepackage{array} 
\usepackage{multirow}

\newcolumntype{x}[1]{%
>{\centering\hspace{0pt}}p{#1}}%
\providecommand{\openone}{\leavevmode\hbox{\small1\kern-3.8pt\normalsize1}}

\def\ie{{\frenchspacing\it i.e.}}
\def\eg{{\frenchspacing\it e.g.}}
\def\etc{{\frenchspacing\it etc.}}

\def\spose#1{\hbox to 0pt{#1\hss}}
\def\simlt{\mathrel{\spose{\lower 3pt\hbox{$\mathchar"218$}}
   \raise 2.0pt\hbox{$\mathchar"13C$}}}
\def\simgt{\mathrel{\spose{\lower 3pt\hbox{$\mathchar"218$}}
     \raise 2.0pt\hbox{$\mathchar"13E$}}}
 \def\simpropto{\mathrel{\spose{\lower 3pt\hbox{$\mathchar"218$}}
     \raise 2.0pt\hbox{$\propto$}}}

\begin{document}
\input{epsf.sty}

% PAPER TEXT

\title{Non-Gaussianity in Two-Field Inflation}

\def\harvard{1}
\def\mit{2}
\def\affilmrk#1{$^{#1}$}

\author{
Courtney M. Peterson\affilmrk{\harvard},
Max Tegmark\affilmrk{\mit}
}

\address{
\parshape 1 -3cm 24cm
\affilmrk{\harvard} Dept.~of Physics, Harvard University, Cambridge, MA 02138, USA \\
\affilmrk{\mit} Dept.~of Physics \& MIT Kavli Institute, Massachusetts Institute of Technology, Cambridge, MA 02139
}

\date{Submitted to Journal Month day year, revised Month day, accepted Month day}
%\date{\today}
\date{Updated: December 14, 2010}

\vspace{10mm}

\begin{abstract}
We derive semi-analytic formulae for the local bispectrum and trispectrum in general two-field inflation
and provide a simple geometric recipe for building observationally allowed models with observable non-Gaussianity.
We use the $\delta N$ formalism and the transfer function formalism to express the bispectrum almost entirely in terms of model-independent physical quantities.
Similarly, we calculate the trispectrum and show that the trispectrum parameter $\tau_{NL}$ can be expressed entirely in terms of spectral observables, which provides a new consistency relation unique to two-field inflation.  
We show that in order to generate observably large non-Gaussianity during inflation, the sourcing of curvature modes by isocurvature modes must be extremely sensitive to the initial conditions, and that the amount of sourcing must be moderate in order to avoid excessive fine-tuning.
Under some minimal assumptions, we argue that the first condition is satisfied only when neighboring trajectories through the two-dimensional field space diverge during inflation.  Geometrically, this means that the inflaton must roll along a ridge in the potential $V$ for some time during inflation and that its trajectory must turn slightly (but not too sharply) in field space.  Therefore, it follows that two-field scenarios with attractor solutions necessarily produce small non-Gaussianity.  
This explains why it has been so difficult to achieve large non-Gaussianity in two-field inflation, and why it has only been achieved in a narrow class of models like hybrid inflation and certain product potentials where the potential and/or the initial conditions are fine-tuned. 
Some of our conclusions generalize qualitatively to general multi-field inflation.
\end{abstract}

\maketitle

\section{Introduction}

Cosmological inflation \cite{Guth-1981,Linde-1990,LythAndRiotto-1998,LiddleAndLyth-2000,BassettEtAl-2005} is widely thought to be responsible for producing the density perturbations that initiated the formation of large-scale structure.  During such an inflationary expansion, quantum fluctuations would have been stretched outside the causal horizon and then frozen in as classical perturbations.  These primordial perturbations would later be 
gravitationally amplified over time into the cosmological large-scale structure that we observe today  
\cite{MukhanovAndChibisov-1981,GuthAndPi-1982,MukhanovAndChibisov-1982,Hawking-1982,Starobinksy-1982,BardeenEtAl-1983}.

Pinning down the specific nature of inflation or whatever physics seeded the primordial density fluctuations is one of the greatest open problems in cosmology.  The simplest models of inflation are driven by a single scalar field producing fluctuations that are adiabatic, nearly scale-invariant, and nearly Gaussian, but these assumptions need to be tested.  Whether the primordial fluctuations were indeed adiabatic and near scale-invariant can be determined by measuring the power spectra of fluctuations;  the upper limit on the isocurvature spectrum constrains non-adiabaticity, while the slope of the scalar (curvature) power spectrum constrains the deviation from scale-invariance.  Similarly, whether the primordial fluctuations obey Gaussian statistics can be tested by measuring reduced $n$-point correlation functions, where $n \ge 3$, since for Gaussian fluctuations, these higher-point functions all vanish and only the two-point function (the power spectrum) is non-zero.  Any deviations from adiabaticity, scale-invariance, and Gaussianity would signal some non-minimal modifications to the simplest scenarios, and hence would provide exciting insight into ultra high energy physics.  

Of these observational measures, non-Gaussianity has the potential to be the most discriminating probe, given all the information contained in higher-point statistics.  This is particularly valuable given how challenging it has been to discriminate among the myriad different inflationary models.   

The two lowest order non-Gaussian measures are the bispectrum and the trispectrum.
Just like the power spectrum $\mathcal{P}_{\mathcal{R}}$ represents the two-point function of the comoving curvature perturbation $\mathcal{R}$ in Fourier space, 
the bispectrum $\mathcal{B}_{\mathcal{R}}$ represents the three-point function and 
the trispectrum $\mathcal{T}_{\mathcal{R}}$ represents the four-point function:
\begin{align}
\label{powerspec eq}
\langle \mathcal{R}(\mathbf{k}_1) \mathcal{R}(\mathbf{k}_2)\rangle = (2\pi)^3 \delta \left(\sum_{i=1}^2\mathbf{k}_i\right) \mathcal{P}_{\mathcal{R}}(\mathbf{k}_1,\mathbf{k}_2),
\end{align}
\begin{align}
\label{bispec eq}
\langle \mathcal{R}(\mathbf{k}_1) \mathcal{R}(\mathbf{k}_2) \mathcal{R}(\mathbf{k}_3) \rangle = (2\pi)^3 \delta \left(\sum_{i=1}^4\mathbf{k}_i\right) \mathcal{B}_{\mathcal{R}}(\mathbf{k}_1,\mathbf{k}_2,\mathbf{k}_3),
% \mathcal{B}_{\mathcal{R}}(\mathbf{k}_1,\mathbf{k}_2,\mathbf{k}_3)^{local} = (2\pi)^4 \left(-\frac{6}{5} f_{NL}^{local} \mathcal{P}_{mathcal{R}\right) \frac{\sum_i^3 k_i^3}{\prod_i^3 k_i^3}
\end{align}
\begin{align}
\label{trispec eq}
\langle \mathcal{R}(\mathbf{k}_1) \mathcal{R}(\mathbf{k}_2) \mathcal{R}(\mathbf{k}_3) \mathcal{R}(\mathbf{k}_4) \rangle = \nonumber\\
=(2\pi)^3 \delta \left(\sum_{i=1}^4\mathbf{k}_i\right)  & \times \mathcal{T}_{\mathcal{R}}(\mathbf{k}_1,\mathbf{k}_2,\mathbf{k}_3,\mathbf{k}_4).
\end{align}
The $\delta$-functions in equations (\ref{powerspec eq})-(\ref{trispec eq}) reflect the fact that the statistical properties are translationally invariant in real space, which makes the above three correlations vanish except if all $\mathbf{k}$-vectors add up to zero --- \ie, are the negative of one another for the power spectrum, form the three sides of a triangle for the bispectrum, and form the sides of a (perhaps non-flat) quadrangle for the trispectrum.
Since the statistical properties are also rotationally invariant, the  power spectrum depends only on the length and not the direction of its vector argument, and the bispectrum depends only on the lengths of the three triangle sides, so that they can be written
simply as $\mathcal{P}_{\mathcal{R}}(k)$ and $\mathcal{B}_{\mathcal{R}}(k_1,k_2,k_3)$, respectively.  

For the bispectrum, it is standard in the literature to define a dimensionless quantity $f_{NL}(k_1,k_2,k_3)$ to represent the bispectrum by dividing by appropriate powers of the power spectrum \cite{Maldacena-2002}\footnote{The factor of $-\frac{6}{5}$ arises from the fact that $f_{NL}$ was originally defined with respect to the metric perturbation.  After inflation ends and during the matter-dominated era, $2\Phi = - \frac{6}{5} \mathcal{R}$.}:
\begin{align}
\label{fNL in terms of bispec}
- \frac{6}{5} f_{NL} (k_1,k_2,k_3) = {B_{\mathcal{R}}(k_1,k_2,k_3)\over  \left[\mathcal{P}_{\mathcal{R}}(k_1) \mathcal{P}_{\mathcal{R}}(k_2) + \mathrm{cyclic \atop permutations}\right]}.
\end{align}
Although $f_{NL}$ can in principle depend on the triangle shape of the $\mathbf{k}$-vectors in very complicated ways, it has been shown that in practice, essentially all models produce an $f_{NL}$ that is well approximated by one of merely a handful of particular functions of triangle shape, with names like ``local'', ``equilateral'', ``warm'',  and ``flat'' 
\cite{BabichEtAl-2004,FergussonAndShellard-2008}.  
% FergussonAndShellard give 4 or 5 depending on how you count
For example, the local function peaks around triangles that are degenerate (with one angle close to zero, like  
for $k_3 \ll k_1 \approx k_2$), while the equilateral function peaks around triangles that are equilateral $(k_1=k_2=k_3)$.
Bispectra dominated by different triangle shapes correspond to different inflationary scenarios and different physics.
In particular, for multi-field inflation, barring non-canonical kinetic terms or higher-order derivative terms in the Lagrangian, the dominant type of bispectra is of the local form \cite{BabichEtAl-2004,FergussonAndShellard-2008}.  Local non-Gaussianity arises from the non-linear evolution of density perturbations once the field fluctuations are stretched beyond the causal horizon.  Seven-year data from WMAP constrains non-Gaussianity of the local form to  \cite{KomatsuEtAl-2010} 
\begin{align}
\label{fNL constraints}
-10 < f_{NL}^{local} < 74   \, \, \, \, (95 \% \, \mathrm{C. L.}),
\end{align}
and a perfect CMB measurement has the potential to detect a bispectrum as low as $|f_{NL}| \approx 3$ \cite{KomatsuAndSpergel-2000}.  

Similarly, the dominant form of trispectra for standard multi-field inflation is also of the local form, and it can be characterized by two dimensionless non-linear parameters, $\tau_{NL}$ and $g_{NL}$.  Five-year data from WMAP constrains these two parameters to \cite{SmidtEtAl-2010}
\begin{align}
\label{trispec constraints}
- 0.6 \times 10^4 < \tau_{NL} < 3.3 \times 10^4   \, \, \, \, (95 \% \, \mathrm{C. L.}), \\
-7.4 \times 10^5 < g_{NL} < 8.2 \times 10^5   \, \, \, \, (95 \% \, \mathrm{C. L.}).
\end{align}

Interestingly, for single-field inflation, the non-linear parameters representing the bispectrum \cite{AllenEtAl-1987,GanguiEtAl-1994,WangAndKamionkowski-2000,AcquavivaEtAl-2002,Maldacena-2002,Creminelli-2003,Gruzinov-2004,CreminelliAndZaldarriaga-2004,SeeryAndLidsey-2005} and trispectrum \cite{OkamotoAndHu-2002,SeeryEtAl-2006,SeeryAndLidsey-2006} are all of order the slow-roll parameters (\ie, at the percent level) and will not be accessible to CMB experiments.  However, if inflation is described by some non-minimal modification, such as multiple fields or higher derivative operators in the inflationary Lagrangian, then non-Gaussianity might be observable in the near future.  Indeed, there have been many attempts to calculate the level of non-Gaussianity in general multi-field models (e.g., \cite{RigopoulosEtAl-2005,RigopoulosEtAl-2005b,KimAndLiddle-2006,BattefeldAndEasther-2006,BattefeldAndBattefeld-2007,YokoyamaEtAl-2007,YokoyamaEtAl-2007b,Huang-2009,ByrnesAndChoi-2010,TanakaEtAl-2010,KimEtAl-2010}), as well as in two-field models (e.g., \cite{BartoloEtAl-2001,BernardeauAndUzan-2002,BernardeauAndUzan-2002b,VernizziAndWands-2006,ChoiEtAl-2007,ByrnesEtAl-2008,Wang-2010,VincentAndCline-2008,MeyersAndSivanandam-2010}). However, it has been very difficult to find models that produce large non-Gaussianity, though some exceptions have been found such as in the curvaton model \cite{LindeAndMukhanov-1996,LythEtAl-2003,BartoloEtAl-2003,LindeAndMukhanov-2005,MalikAndLyth-2006,SasakiEtAl-2006}, hybrid and multi-brid inflation (e.g., \cite{EnqvistAndVaihkonen-2004,AlabidiAndLyth-2006,Alabidi-2006,BarnabyAndCline-2006,BarnabyAndCline-2006b,ByrnesEtAl-2008b,Sasaki-2008,NarukoAndSasaki-2008,Huang-2009}), and in certain modulated and tachyonic (p)re-heating scenarios (e.g., \cite{DvaliEtAl-2004,Zaldarriaga-2003,SuyamaAndYamaguchi-2007,IchikawaEtAl-2008,EnqvistEtAl-2004,EnqvistEtAl-2005,JokinenAndMazumdar-2005}).  Moreover, it has not been wholly clear why it is so difficult to produce large non-Gaussianity in such models.  Though some authors \cite{BernardeauAndUzan-2002,BernardeauAndUzan-2002b,RigopoulosEtAl-2005,RigopoulosEtAl-2005b,VernizziAndWands-2006,TanakaEtAl-2010} have found spikes in non-Gaussianity whenever the inflaton trajectory changes direction sharply, these spikes in non-Gaussianity are transient and die away before the end of inflation. That makes a comprehensive study of non-Gaussianity generation timely, to understand any circumstances under which observably large non-Gaussianity arises in such models.

In this paper, we calculate the bispectrum and trispectrum in general inflation models with standard kinetic terms, focusing on the important case of two-field inflation. We provide conditions for large non-Gaussianity and a unified answer to the mystery of why it has been so hard to produce large non-Gaussianity in two-field inflationary models.  The rest of this paper is organized as follows.  In Section \ref{background eqtns}, we present the background equations of motion for the fields and discuss the field vector kinematics.  Section \ref{transfer funcs & spec} presents the equations of motion for the field perturbations and some necessary results for the power spectra.  In Section \ref{bispec}, we describe the $\delta N$ formalism, which we use to calculate the bispectrum.  In tandem, we discuss the necessary conditions for large non-Gaussianity.  Finally, we tackle the trispectrum in Section \ref{trispec}.  We summarize our conclusions in Section \ref{conclusions}.

\section{Background Field Equation \& Kinematics} 
\label{background eqtns}

In this section, we review the background equations of motion and discuss the kinematics of the background fields.  This discussion will help us calculate the primordial bispectrum and trispectrum in two-field inflation and understand what features are necessary for non-Gaussianity to be observably large.

We consider general two-field inflation where the non-gravitational part of the action is of the form
\begin{align}
\label{action}
S = \int \left[-\frac{1}{2} g^{\mu \nu} \delta_{ij} \frac{\partial \phi^i}{\partial x^{\mu}} \frac{\partial \phi^j}{\partial x^{\nu}} - V(\phi_1,\phi_2)\right] \sqrt{-g} \, d^4x,
\end{align}
where $V(\phi_1,\phi_2)$ is a completely arbitrary potential of the two fields, $g_{\mu \nu}$ is the spacetime metric, and $\delta_{ij}$ reflects the fact that we assume the kinetic terms are canonical.

In this paper, we adopt similar notation to \cite{PetersonAndTegmark-2010}: boldface for vectors, the symbol $^{T}$ to denote the transpose of a vector, and standard vector product notation.  Gradients and partial derivatives represent derivatives with respect to the fields, except where explicitly indicated otherwise.  We set the reduced Planck mass, $\bar{m} \equiv 8\pi G$, equal to unity, so that all fields are measured in units of the reduced Planck mass.  To simplify the equations of motion and connect them more directly with observables, we use the number of $e$-folds, $N$, as our time variable. $N$ is defined through the relation
\begin{eqnarray}
dN = H dt,
\end{eqnarray}
where $t$ is the comoving time and $H$ is the Hubble parameter.  We denote derivatives with respect to $N$ using the notation
\begin{eqnarray}
' = \frac{d}{dN}.
\end{eqnarray}

Using $N$ as the time variable, we showed in \cite{PetersonAndTegmark-2010} that the background equation of motion for the fields can be written as
\begin{eqnarray}
\label{Phi EoM}
\frac{\boldsymbol{\eta}}{(3 - \epsilon)} + \boldsymbol{\phi}' = - \boldsymbol{\nabla} \ln V.
\end{eqnarray}
The parameter $\epsilon$ is defined as
\begin{align}
\label{epsilon}
\epsilon \equiv -(\ln H)' = \frac{1}{2} \boldsymbol{\phi}' \cdot \boldsymbol{\phi}' ,
\end{align}
and $\boldsymbol{\eta}$ is the field acceleration, defined as
\begin{align}
\boldsymbol{\eta} \equiv \boldsymbol{\phi}''.
\end{align}
%where the symbol $\frac{D}{dN}$ acting on a contravariant vector means
%\begin{align}
%\frac{DX^{\mu}}{dN} \equiv \frac{dX^{\mu}}{dN} + \Gamma^{\mu}_{\nu \sigma} X^{\nu} \frac{d\phi^\sigma}{dN},
%\end{align} 
%where $\Gamma^{\mu}_{\nu \sigma}$ is the Christoffel symbol corresponding to the field metric $G_{ij}$. 

In \cite{PetersonAndTegmark-2010}, we also explained how the two quantities $\boldsymbol{\phi}'$ and $\boldsymbol{\eta}$ represent the kinematics of the background fields.  If we view the fields as coordinates on the field manifold, then $\boldsymbol{\phi}'$ represents the field velocity, and 
\begin{align}
v \equiv |\boldsymbol{\phi}'|
\end{align}
represents the field speed.  Similarly, $\boldsymbol{\eta}$ is the field acceleration.

The velocity vector, $\boldsymbol{\phi}'$, is also useful because it can be used to define a {\it kinematical basis} \cite{GordonEtAl-2000,NibbelinkAndVanTent-2000,NibbelinkAndVanTent-2001}.  In this basis, the basis vector $\boldsymbol{e}_{\parallel}$ points along the field trajectory, while the basis vector $\boldsymbol{e}_{\perp}$ points perpendicularly to the field trajectory, in the direction that makes the scalar product $\boldsymbol{e}_{\perp} \cdot \boldsymbol{\eta}$ positive.   To denote the components of a vector and a matrix in this basis, we use the short-hand notation
\begin{align}
\label{vec cmpts in kine basis}
X_{\parallel} \equiv \mathbf{e}_{\parallel} \cdot \mathbf{X}, \, \, \, \, \, \, \, \, \, \, \, \, \, X_{\perp} \equiv \mathbf{e}_{\perp} \cdot \mathbf{X},
\end{align}
and 
\begin{align}
\label{matrix cmpts in kine basis}
M_{\parallel \perp} \equiv \mathbf{e}_{\parallel}^T \, \mathbf{M} \, \mathbf{e}_{\perp}, \, \, \, \, \, \, \, \, \, \, \, \, \, \, \,  \etc 
\end{align}
The kinematical basis is useful for several reasons.  First, the field perturbations naturally decompose into components parallel and perpendicular to the field trajectory, and the former represent \textit{bona fide} density perturbations, while the latter do not.  This decomposition of the field perturbations is helpful in finding expressions for the power spectra.  Second, it allows us to consider separate aspects of the background field kinematics, which in \cite{PetersonAndTegmark-2010}, we encapsulated in a set of three quantities.  The first quantity is the field speed, $v$.  The second and third quantities arise from decomposing the field acceleration into components parallel and perpendicular to the field velocity.   In particular, the quantity $\frac{\eta_\parallel}{v}$ represents the logarithmic rate of change in the field speed (the \textit{speed-up rate}), while the quantity $\frac{\eta_\perp}{v}$  represents  the rate at which the field trajectory changes direction (the \textit{turn rate}) \cite{PetersonAndTegmark-2010}.   

This distinction between the speed-up rate and the turn rate is important for two reasons.  First, the turn rate represents uniquely multi-field behavior (as the background trajectory cannot turn in single-field inflation), whereas the speed-up rate represents single-field-like behavior.  Second, the speed-up and turn rates have very different effects on the evolution of the field perturbations and hence on the power spectra.  Indeed, the features in the power spectra depend not only on the absolute sizes of the two rates but also on their relative sizes to each other; in particular, the ratio of the turn rate to the speed-up rate is an indicator of the relative impact of multi-field effects.  So disentangling the two quantities allows for a better understanding of the power spectra and all the ways that the spectra can be made consistent with observations.

To fully take advantage of this distinction between the speed-up and turn rates, we redefined the standard slow-roll approximation in \cite{PetersonAndTegmark-2010}, splitting it into two different approximations that can be invoked either separately or together.  As background, the standard slow-roll approximation is typically expressed as
\begin{align}
\label{epsilon in SRST}
\epsilon \approx \frac{1}{2} |\boldsymbol{\nabla} \ln V|^2 \ll 1,
\end{align}
and 
\begin{align}
\label{2nd cond of slow-roll}
\left|\frac{\partial_i \partial_j V}{V} \right| \ll 1.
\end{align}
However, as argued in \cite{PetersonAndTegmark-2010}, the latter condition lumps together and simultaneously forces the speed-up rate, the turn rate, and a quantity called the entropy mass to be small.  So instead, we redefined the slow-roll approximation to mean that the field speed is small, 
\begin{align}
\epsilon = \frac{1}{2} v^2 \ll 1,
\end{align}
and is slowly changing,
\begin{align}
\frac{\eta_{\parallel}}{v} \ll 1.
\end{align}
In other words, the above slow-roll approximation represents the minimum conditions necessary to guarantee quasi-exponential inflationary expansion and corresponds to limits on single-field-like behavior.  As for the turn rate, we endowed it with its own separate approximation, the {\it slow-turn approximation}, which applies when the turn rate satisfies
\begin{align}
\frac{\eta_{\perp}}{v} \ll 1.
\end{align}
The slow-turn limit corresponds to limits on multi-field behavior.  Finally, this alternative framework allows the lowest order entropy mass to take on any arbitrary value. (See \cite{PetersonAndTegmark-2010} for further discussion of these points.)  

When the background field vector is both slowly rolling and slowly turning, we call the combined slow-roll and slow-turn limits (which is equivalent to the conventional slow-roll limit minus the constraint on the entropy mass) the SRST limit for brevity.  In the combined limit, the evolution equation for the fields can be approximated by 
\begin{eqnarray}
\label{Phi EoM in SRST}
\boldsymbol{\phi}' = - \boldsymbol{\nabla} \ln V.
\end{eqnarray}
Also in this combined limit, the speed-up rate and the turn rate can be approximated by 
\begin{align}
\frac{\eta_{\parallel}}{v} \approx - M_{\parallel \parallel}, \, \, \, \, \, \, \, \, \, \, \, \, \, \, \, \, \, \, \, \, \frac{\eta_{\perp}}{v} \approx - M_{\parallel \perp},
\end{align}
respectively, where we define the {\it mass matrix}, $\mathbf{M}$, as the Hessian of $\ln V$, \ie,
\begin{align}
\label{M defn}
\mathbf{M} \equiv \boldsymbol{\nabla} \boldsymbol{\nabla}^T \ln V.
\end{align}
Being a symmetric $2\times 2$ matrix, $\mathbf{M}$ is characterized by three independent coefficients.  In the kinematical basis, these three coefficients are
$M_{\parallel\parallel}$, $M_{\parallel\perp}$, and $M_{\perp\perp}$, the third of which is the lowest order {\it entropy mass}.\footnote{We refer to $M_{\perp\perp}$ as the entropy {\it mass}, even though we constructed it to be dimensionless.}
In other words, in the kinematical basis and under the SRST limit, we can interpret the mass matrix  as follows:
%\begin{align}
%{\bf M} 
%& = 
%\left(
%\begin{tabular}{cc}
%$M_{\parallel\parallel}$		&$M_{\parallel\perp}$\\
%$M_{\parallel\perp}$		&$M_{\perp\perp}$
%\end{tabular}
%\right)
%\\ & = 
%\left(
%\begin{tabular}{cc}
%{\footnotesize $-$speed-up rate}		&{\footnotesize $-$turn rate}\\
%{\footnotesize $-$turn rate}		&{\footnotesize entropy mass}
%\end{tabular}
%\right),
%\nonumber
%\end{align}
\begin{align}
{\bf M} 
 = 
\left(
\begin{tabular}{cc}
$M_{\parallel\parallel}$		&$M_{\parallel\perp}$\\
$M_{\parallel\perp}$		&$M_{\perp\perp}$
\end{tabular}
\right)
 = 
\left(
\begin{tabular}{cc}
{\footnotesize $-$speed-up rate}		&{\footnotesize $-$turn rate}\\
{\footnotesize $-$turn rate}		&{\footnotesize entropy mass}
\end{tabular}
\right),
\end{align}
where the speed-up rate and turn rate alone determine the background kinematics.  However, all three quantities --- the speed-up rate, the turn rate, and the entropy mass --- affect how the perturbations evolve, as described in the next section.

\section{Perturbations, Transfer Functions, and Power Spectra}
\label{transfer funcs & spec}

In this section, we summarize the general results for the evolution of perturbations and for the power spectra, as these expressions will enable us to calculate the bispectrum and trispectrum in two-field inflation and to express the results in terms of other spectral observables.

In \cite{PetersonAndTegmark-2010}, we derived the following general equation of motion for the field perturbations in Fourier space:
\begin{align}
\label{Delta Phi EoM wrt N}
\frac{1}{(3 - \epsilon)} \frac{D^2\boldsymbol{\delta\phi}}{dN^2} + \frac{D\boldsymbol{\delta\phi}}{dN} & + \left(\frac{k^2}{a^2 V}\right) \boldsymbol{\delta\phi} \nonumber \\ & = - \left[\mathbf{M} + \frac{\boldsymbol{\eta}\boldsymbol{\eta}^T}{(3-\epsilon)^2}\right] \boldsymbol{\delta\phi},
\end{align}
where $\boldsymbol{\delta\phi}$ represents the field perturbation in the flat gauge, which coincides with the gauge-invariant Mukhanov-Sasaki variable \cite{Sasaki-1986,Mukhanov-1988}.  For modes in the super-horizon limit ($k\ll aH$), we showed that when the background fields are in the SRST limit, the above expression reduces to \cite{PetersonAndTegmark-2010}
\begin{eqnarray}
\label{Delta Phi EoM wrt N in SRST}
\frac{D\boldsymbol{\delta\phi}}{dN} \approx - \mathbf{M} \, \boldsymbol{\delta\phi}.
\end{eqnarray}

Now we switch to working in the kinematical basis, where the modes decompose into adiabatic modes, $\delta \phi_{\parallel}$, and entropy modes, $\delta \phi_{\perp}$.  The former represent density perturbations, while the latter represent relative field perturbations that leave the overall density unperturbed.   In this basis, the super-horizon equations of motion for the two mode types are
\begin{align}
\label{mode EoMs}
\delta \phi_{\parallel}' & = \left(\frac{\eta_\parallel}{v}\right) \delta \phi_{\parallel} + 2 \left(\frac{\eta_\perp}{v}\right) \delta \phi_{\perp}, \nonumber \\
\delta \phi_{\perp}'  & \approx - M_{\perp \perp} \delta \phi_{\perp}, 
\end{align}
where the first equation is exact and the second is valid to lowest order in the slow-turn limit.  (Full expressions are given in \cite{PetersonAndTegmark-2010}.)   In the SRST limit, the evolution of modes is determined by the three unique coefficients of the mass matrix.  The evolution of adiabatic modes is controlled by $\frac{\eta_{\parallel}}{v} \approx - M_{\parallel \parallel}$ and by $\frac{\eta_{\perp}}{v} \approx - M_{\parallel \perp}$.   The third unique coefficient of the mass matrix, $M_{\perp \perp}$, alone determines the relative damping or growth of entropy modes.  We call $M_{\perp \perp}$ the lowest order entropy mass (or just the entropy mass) because it approximates the effective mass in the full second-order differential equation of motion for the entropy modes \cite{PetersonAndTegmark-2010}.  In addition to being viewed as an effective mass, $M_{\perp \perp}$ can also be viewed as a measure of the curvature of the potential along the $\mathbf{e}_{\perp}$ or entropic direction.  When the curvature of the potential along the entropic direction is positive, the entropy modes decay; when the curvature is negative, the entropy modes grow.

Directly related to these two modes are the curvature and isocurvature modes, the two quantities whose power spectra are typically computed when considering the two-field power spectra.  During inflation, the curvature and isocurvature modes are simply related to the adiabatic and entropy modes, respectively, by a factor of $\frac{1}{v}$ \cite{WandsEtAl-2002}.  That is, the curvature modes are given by
\begin{align}
\label{comoving R in terms of adiabatic perturbation}
\mathcal{R} = \frac{\delta \phi_{\parallel}}{v},
\end{align}  
and the isocurvature modes by 
\begin{align}
\label{isocurv defn}
\mathcal{S} \equiv \frac{\delta \phi_{\perp}}{v}. 
\end{align} 
The super-horizon evolution of curvature and isocurvature modes can be determined from the equations of motion for the adiabatic and entropy modes.  We parametrize the solutions through the transfer matrix formalism \cite{AmendolaEtAl-2001,WandsEtAl-2002}:
\begin{align}
\label{transfer matrix}
\left(\begin{array}{c} \mathcal{R} \\ \mathcal{S} \end{array} \right) = & 
\left(\begin{array}{cc} 1 & T_{\mathcal{RS}} \\ 0 & T_{\mathcal{SS}} \end{array} \right)
\left(\begin{array}{c} \mathcal{R}_* \\ \mathcal{S}_* \end{array} \right),
\end{align}
where the transfer functions can be written as
\begin{align}
\label{transfer functions}
T_{\mathcal{RS}}(N_*,N) & \equiv \int_{N_*}^N \alpha(\tilde{N}) \, T_{\mathcal{SS}}(N_*,\tilde{N}) \, d\tilde{N}, \nonumber \\
T_{\mathcal{SS}}(N_*,N) & \equiv e^{\int_{N_*}^N \beta(\tilde{N}) \,  d\tilde{N}}.
\end{align}
The subscript $*$ means that the quantity is to be evaluated when the corresponding modes exit the horizon.  The transfer function $T_{\mathcal{SS}}$ therefore represents how much the isocurvature modes have decayed (or grown) after exiting the horizon. The transfer function $T_{\mathcal{RS}}$ represents the total sourcing of curvature modes by isocurvature modes; that is, it represents the importance of the multi-field effects.  In \cite{PetersonAndTegmark-2010}, we found that 
\begin{align}
\label{alpha}
\alpha & = 2 \frac{\eta_{\perp}}{v} 
\end{align}
exactly, which tells us that the curvature modes are only sourced by the isocurvature modes when the field trajectory changes direction.  However, the \textit{isocurvature mass}, $\beta$, must be approximated or computed numerically.  To lowest order in the SRST limit,
\begin{align}
\label{alpha and beta in SRST}
\alpha & = -2 M_{\parallel \perp}, \nonumber \\
\beta & = M_{\parallel \parallel} - M_{\perp \perp}.
\end{align}
From a geometrical perspective, equation (\ref{alpha and beta in SRST}) for $\beta$ shows that how fast the isocurvature modes evolve depends on the difference between the curvatures of the potential along the entropic and adiabatic directions.  From a kinematical perspective, the isocurvature modes will grow if the entropy modes grow faster than the field vector picks up speed. This means that isocurvature modes tend to grow in two types of scenarios: when $M_{\perp \perp}$ is  large and negative and when $\epsilon$ decreases quickly.  Otherwise, when the entropy modes do not grow faster than the field vector picks up speed, the isocurvature modes decay.

Now we present expressions for the power spectra and their associated observables.  The power spectrum of a quantity $\mathcal{X}$ is essentially the variance of its  Fourier transform:
\begin{align}
\langle \mathcal{X}(\mathbf{k}_1) \mathcal{X}(\mathbf{k}_2) \rangle = (2\pi)^3 \delta (\mathbf{k}_1 + \mathbf{k}_2) \mathcal{P}_{\mathcal{X}}(k_1).
\end{align}
Using the above results, the curvature, cross, and isocurvature spectra at the end of inflation can be written to lowest order as \cite{WandsEtAl-2002}
\begin{align}
\label{pwr spec}
\mathcal{P}_{\mathcal{R}} & = \left(\frac{H_*}{2\pi}\right)^2 \frac{1}{2\epsilon_*} (1 + T_{\mathcal{RS}}^2), \nonumber \\
\mathcal{C}_{\mathcal{RS}} & = \left(\frac{H_*}{2\pi}\right)^2 \frac{1}{2\epsilon_*}  T_{\mathcal{RS}} T_{\mathcal{SS}}, \\
\mathcal{P}_{\mathcal{S}} & = \left(\frac{H_*}{2\pi}\right)^2 \frac{1}{2\epsilon_*}  T_{\mathcal{SS}}^2, \nonumber
\end{align}
where it is implied that the transfer functions are evaluated at the end of inflation.  The associated curvature spectral index is \cite{PetersonAndTegmark-2010}
\begin{align}
\label{nR}
n_{\mathcal{R}}\equiv n_s-1 \equiv \frac{d \ln P_{\mathcal{R}}}{dN} = \left[n_T + 2 \mathbf{e}_N^T \mathbf{M} \mathbf{e}_N\right]_*, 
\end{align}
where $n_s$ is the standard scalar spectral index that is constrained by observations, $n_T = -2\epsilon_*$ is the tensor spectral index, and the unit vector $\mathbf{e}_N$ points in the direction of the gradient of $N$.  In the kinematical basis, $\mathbf{e}_N$ takes the form \cite{PetersonAndTegmark-2010}
\begin{align}
\label{e_N in kine basis}
\mathbf{e}_N = \cos \Delta_N \, \mathbf{e}_{\parallel}^* + \sin \Delta_N \, \mathbf{e}_{\perp}^*,
\end{align}
where $\Delta_N$ is the correlation angle, which satisfies 
\begin{align}
\label{tan Delta N}
\tan \Delta_N = T_{\mathcal{RS}}.
\end{align}
The correlation angle can be given in terms of the dimensionless curvature-isocurvature correlation, $r_C$, which we define as \cite{PetersonAndTegmark-2010}
\begin{align}
\label{rC}
r_C \equiv \frac{C_{\mathcal{RS}}}{\sqrt{P_{\mathcal{R}} P_{\mathcal{S}}}} = \sin \Delta_N,
\end{align}
in analogy to the tensor-to-scalar ratio,
\begin{align}
r_T \equiv \frac{\mathcal{P}_T}{\mathcal{P}_{\mathcal{R}}} = 16 \epsilon_* \cos^2 \Delta_N,
\end{align}
where $\mathcal{P}_T$ is the tensor spectrum of gravitational waves.
Similarly, another ratio of spectra, the isocurvature fraction, 
\begin{align}
f_{iso} \equiv \frac{P_\mathcal{S}}{P_{\mathcal{R}}} = \frac{T_{\mathcal{SS}}^2}{1+T_{\mathcal{RS}}^2} = \cos^2 \Delta_N T_{\mathcal{SS}}^2,
\end{align}
gives the relative size of $T_{\mathcal{SS}}$ to $T_{\mathcal{RS}}$.

These results provide the starting point for our computation of the bispectrum and trispectrum for two-field inflation.

\section{The Bispectrum}
\label{bispec}

In this section, we calculate the bispectrum for general two-field inflation using the $\delta N$ formalism and the transfer function formalism.  We also develop a set of conditions encapsulating when the bispectrum is large enough to be detected.

\subsection{Calculation of $f_{NL}$ Using the $\delta N$ Formalism}
\label{bispec initial calc}

We focus on bispectrum configurations of the local or squeezed type (\eg, $k_3 \ll k_1 \approx k_2$), which is the dominant type present during standard multi-field inflation.   As we showed earlier in equation (\ref{fNL in terms of bispec}), the bispectrum can be expressed in terms of the dimensionless non-linear parameter $f_{NL}$ \cite{Maldacena-2002}.\footnote{The non-linear parameter $f_{NL}$ was originally introduced to represent the degree of non-Gaussianity in the metric perturbation \cite{VerdeEtAl-1999,KomatsuAndSpergel-2000},
\begin{align}
\label{defn of fNL}
\Phi = \Phi_G + f_{NL} \Phi_G^2,
\end{align} 
where $\Phi_G$ is Gaussian and $\Phi$ is not. Here,
$\Phi$ is the metric perturbation in the Newtonian gauge, which equals the gauge-invariant Bardeen variable.}  From here on, whenever $f_{NL}$ appears in this paper, it represents the local form, so we drop the superscript \textit{local}.

Conveniently, $f_{NL}$ can be written in terms of the $\delta N$ formalism \cite{Starobinsky-1985,SasakiAndStewart-1995,LythEtAl-2004}, where $N$ again represents the number of $e$-folds of inflation.  Under the $\delta N$ formalism, $\mathcal{R} = \boldsymbol{\nabla} N \cdot \delta \boldsymbol{\phi}$ \cite{SasakiAndStewart-1995,LythEtAl-2004}, where $\delta \boldsymbol{\phi}$ is measured in the flat gauge and where it is implied that the gradient is with respect to the fields at horizon exit. (For brevity, we drop the subscript $*$ on $\boldsymbol{\nabla}$ in this section, but restore it in later sections whenever there might be some potential ambiguity.)  Using this result, correlators of $\mathcal{R}$ can be written in terms of gradients of $N$.  In particular, it has been shown \cite{LythAndRodriguez-2005} that the local form of $f_{NL}$ can written as 
\begin{align}
\label{fNL4 delta N formula}
-\frac{6}{5} f_{NL} & = \frac{\boldsymbol{\nabla}^T N \, \boldsymbol{\nabla} \boldsymbol{\nabla}^T N \, \boldsymbol{\nabla} N}{|\boldsymbol{\nabla} N|^4}.
\end{align}

We use equation (\ref{fNL4 delta N formula}) to find an expression for $f_{NL}$ in terms of the transfer function formalism.  We start by first finding a semi-analytic formula for $\boldsymbol{\nabla} N$ in two-field inflation.  By comparing equation (\ref{pwr spec}) to the lowest order result for the curvature power spectrum in multi-field inflation \cite{SasakiAndStewart-1995},
\begin{align}
\label{spec in terms of grad N}
\mathcal{P}_{\mathcal{R}} = \left(\frac{H_*}{2\pi}\right)^2 |\boldsymbol{\nabla} N|^2,
\end{align}
we obtain
\begin{align}
\label{norm grad N}
|\boldsymbol{\nabla} N| = \sqrt{\frac{1+T_{\mathcal{RS}}^2}{2\epsilon_*}},
\end{align}
where again it is implied that $T_{\mathcal{RS}}$ is evaluated at the end of inflation.  Combining equation (\ref{norm grad N}) with the fact that
\begin{align}
\boldsymbol{\phi}' \cdot \boldsymbol{\nabla} N = 1,
\end{align}
we conclude that $\boldsymbol{\nabla} N$ takes the following form in the kinematical basis:
\begin{align}
\label{nabla N formula}
\boldsymbol{\nabla} N = \frac{1}{\sqrt{2\epsilon_*}} \left(\boldsymbol{e}_{\parallel}^* +  T_{\mathcal{RS}} \boldsymbol{e}_{\perp}^*\right).
\end{align}
The above equation implies that we can also write  $\boldsymbol{\nabla} N$ as
\begin{align}
\label{nabla N formula v2}
\boldsymbol{\nabla} N 
= \sqrt{\frac{1+T_{\mathcal{RS}}^2}{2\epsilon_*}} \, \mathbf{e}_N 
=  \frac{\mathbf{e}_N}{\sqrt{2\epsilon_*} \cos \Delta_N} \,  ,
\end{align}
where $\mathbf{e}_N$ is the unit vector in the direction of $\boldsymbol{\nabla} N$ and is given by equation (\ref{e_N in kine basis}).

Next, we re-write equation (\ref{fNL4 delta N formula}) for $f_{NL}$ as
\begin{align}
\label{f_NL^4 delta N formula v2}
-\frac{6}{5}f_{NL} & = \frac{\boldsymbol{e}_N^T \, \boldsymbol{\nabla} \boldsymbol{\nabla}^T N \, \boldsymbol{e}_N}{|\boldsymbol{\nabla} N|^2}.
\end{align}
Since $\boldsymbol{e}_N \cdot \boldsymbol{e}_N = 1$, it follows that $\boldsymbol{\nabla} \boldsymbol{e}_{N} \cdot \boldsymbol{e}_{N} = 0$, and hence
\begin{align}
\boldsymbol{\nabla} \boldsymbol{\nabla}^T N \, \mathbf{e}_N =  \boldsymbol{\nabla} |\boldsymbol{\nabla} N|.
\end{align}  
Taking this result, dividing through by $|\boldsymbol{\nabla} N|$ and using equations (\ref{tan Delta N}) and (\ref{norm grad N}), we find
\begin{align}
\label{grad of grad N}
\frac{\boldsymbol{\nabla} \boldsymbol{\nabla}^T N \, \mathbf{e}_N}{|\boldsymbol{\nabla} N|}  =  - \frac{\boldsymbol{\nabla} \epsilon_*}{2\epsilon_*} + \sin \Delta_N \cos \Delta_N \boldsymbol{\nabla} T_{\mathcal{RS}}. 
\end{align}
In the SRST limit, using equations (\ref{epsilon}) and (\ref{Phi EoM in SRST}), it holds that
\begin{align}
\boldsymbol{\nabla} \epsilon = - \boldsymbol{M} \boldsymbol{\phi}' .
\end{align}
Substituting this result into equation~(\ref{grad of grad N}) and dividing through by another factor of $|\boldsymbol{\nabla} N|$, we find that
\begin{align}
\label{key delta N step}
\frac{\boldsymbol{\nabla} \boldsymbol{\nabla}^T N \, \mathbf{e}_N}{|\boldsymbol{\nabla} N|^2}  =  & \cos \Delta_N \times \\ & \left[\boldsymbol{M}^*  \mathbf{e}_{\parallel}^* + \sin \Delta_N \cos \Delta_N \sqrt{2\epsilon_*} \, \boldsymbol{\nabla} T_{\mathcal{RS}}\right]. \nonumber
\end{align}

To complete our calculation of $f_{NL}$, we need to contract equation (\ref{key delta N step}) with the unit vector $\mathbf{e}_N$.  We break this calculation into two parts, based on the fact that $\boldsymbol{e}_N = \cos \Delta_N \, \mathbf{e}_\parallel^* + \sin \Delta_N \, \mathbf{e}_\perp^*$.  First, we contract $\cos \Delta_N (\mathbf{e}_\parallel^*)^T$ with equation (\ref{key delta N step}).  Using $\frac{d}{dN} = \boldsymbol{\phi}' \cdot \boldsymbol{\nabla}$ and the norm of equation (\ref{nabla N formula v2}), we can write
\begin{align}
\frac{\cos \Delta_N \, (\mathbf{e}_{\parallel}^*)^T \, \boldsymbol{\nabla} \boldsymbol{\nabla}^T N \, \mathbf{e}_N}{|\boldsymbol{\nabla} N|^2}  = \cos^2 \Delta_N \frac{d}{dN} \ln |\boldsymbol{\nabla} N|.
\end{align}  
From equations (\ref{nR}) and (\ref{spec in terms of grad N}), it follows that 
\begin{align}
\label{cos Delta para cmpt of key delta N step}
\frac{\cos \Delta_N \, (\mathbf{e}_{\parallel}^*)^T \, \boldsymbol{\nabla} \boldsymbol{\nabla}^T N \, \mathbf{e}_N}{|\boldsymbol{\nabla} N|^2}  = \frac{1}{2} \cos^2 \Delta_N \, (n_{\mathcal{R}}-n_T).
\end{align}  
Second, we calculate $\sin \Delta_N \, (\mathbf{e}_{\perp}^*)^T$ contracted with equation (\ref{key delta N step}), which yields
\begin{align}
\label{sin Delta para cmpt of key delta N step}
\frac{\sin \Delta_N \, (\mathbf{e}_\perp^*)^T \, \boldsymbol{\nabla} \boldsymbol{\nabla}^T N \, \mathbf{e}_N}{|\boldsymbol{\nabla} N|^2}   =  \sin \Delta_N \cos \Delta_N \times \, \, \, \, \, \, \, \, \, \, \, \, \, \, \, \, \\  \, \, \, \, \, \,  \, \, \, \, \, \, \left[M_{\parallel \perp}^* + \sin \Delta_N \cos \Delta_N \sqrt{2\epsilon_*} \, \mathbf{e}_{\perp}^* \cdot \boldsymbol{\nabla} T_{\mathcal{RS}}\right]. \nonumber
\end{align}
Combining equations (\ref{cos Delta para cmpt of key delta N step}) and (\ref{sin Delta para cmpt of key delta N step}), $f_{NL}$ can be written as
\begin{widetext}
\begin{align}
\label{fNL general expression}
-\frac{6}{5}f_{NL} = \frac{1}{2}  (n_{\mathcal{R}} - n_T)\cos^2 \Delta_N + \left(M_{\parallel \perp}^* +  \sin \Delta_N \cos \Delta_N \sqrt{2\epsilon_*} \, \mathbf{e}_{\perp}^* \cdot \boldsymbol{\nabla} T_{\mathcal{RS}} \right) \sin \Delta_N \cos \Delta_N.
\end{align}
\end{widetext}

Equation (\ref{fNL general expression}) depends on sines and/or cosines times the curvature (scalar) spectral index, the tensor spectral index, the turn rate $\frac{\eta_{\perp}}{v} \approx - M_{\parallel \perp}$ at horizon exit, and $\sqrt{2\epsilon}_* \mathbf{e}_{\perp}^* \cdot \boldsymbol{\nabla} T_{\mathcal{RS}}$.  Observational constraints force the magnitudes of $n_{\mathcal{R}}$ and $n_T$ to be much less than unity, and the turn rate at horizon exit must be at least somewhat less than unity to avoid violating scale-invariance and causing a complete breakdown of the SRST approximation at horizon exit.  Therefore, the magnitude of
 $f_{NL}$ cannot exceed unity unless 
\begin{align}
\label{cond on grad TRS}
\boxed{\left| \sin^2 \Delta_N \cos^2 \Delta_N \, \left(\mathbf{e}_{\perp}^* \cdot \boldsymbol{\nabla} T_{\mathcal{RS}}\right)\right| \simgt \frac{1}{\sqrt{2\epsilon_*}}.}
\end{align}
Physically speaking, for equation (\ref{cond on grad TRS}) to be satisfied requires that two conditions be met:
\begin{enumerate}
\item The total amount of sourcing of curvature modes by isocurvature modes ($T_{\mathcal{RS}}$) must be extremely sensitive to a change in the initial conditions perpendicular to the inflaton trajectory.  In other words, two neighboring trajectories must experience dramatically different amounts of sourcing.
\item The total amount of sourcing must be non-zero (\ie, $\sin \Delta_N \neq 0$).  Usually, the amount of sourcing must also be moderate, to avoid having $\sin^2 \Delta_N \cos \Delta_N^2 \lll 1$.  
\end{enumerate}

The first condition, that $T_{\mathcal{RS}}$ be extremely sensitive to the initial conditions, makes sense on an intuitive level.  In order to produce large non-Gaussianity, perturbations off the classical trajectory must move the inflaton onto neighboring trajectories that experience very different dynamics for the field perturbations, hence producing a large degree of skew in the primordial fluctuations.  Meeting this condition that $T_{\mathcal{RS}}$ is very sensitive to the initial conditions can be achieved in a couple of different ways.  Since equations (\ref{transfer functions}) and (\ref{alpha}) show that $T_{\mathcal{RS}}$ depends on an integral of the turn rate times the relative amplitude of isocurvature modes ($T_{\mathcal{SS}}$), \textit{neighboring trajectories need to have very different turn rate profiles, $T_{\mathcal{SS}}$ profiles, or both}. 

The second condition reflects the fact that in the limit of no sourcing --- which corresponds to single-field behavior --- the bound in equation (\ref{cond on grad TRS}) can never be satisfied.  Usually, the sourcing must also be moderate, but this is not a strict requirement per se.  However, if the total sourcing is tiny ($\sin \Delta_N \ll 1$) or is very large ($\cos \Delta_N \ll 1$), then the trigonometric terms 
will usually prevent the bound in equation (\ref{cond on grad TRS}) from being satisfied.   We add that to achieve moderate sourcing, the larger $T_{\mathcal{SS}}$ is during inflation, the smaller the turn rate must be, and vice versa.   For reference, Figure 1 shows the value of the trigonometric factor $\sin^2 \Delta_N \cos \Delta_N^2$ as a function of the total mode sourcing, $T_{\mathcal{RS}}$.  Its maximum value is 0.25, which occurs at $T_{\mathcal{RS}} = 1$.  

\begin{figure}[t]
\includegraphics[height=60mm]{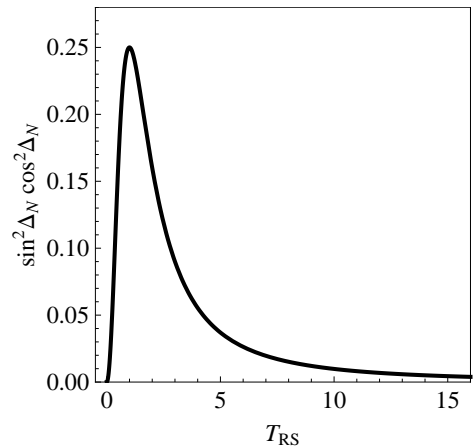}
\label{FigB}
\caption{The trigonometric factor $\sin^2 \Delta_N \cos^2 \Delta_N$ as a function of the total amount of mode sourcing, $T_{\mathcal{RS}}$.  For reference, when $T_{\mathcal{RS}} = 1$, half of the curvature (scalar) power spectrum at the end of inflation is due to the sourcing of curvature modes by isocurvature modes.}
\end{figure}  

\begin{figure*}[t]
\begin{minipage}{1.0\linewidth}
\centering
\includegraphics[height=120mm]{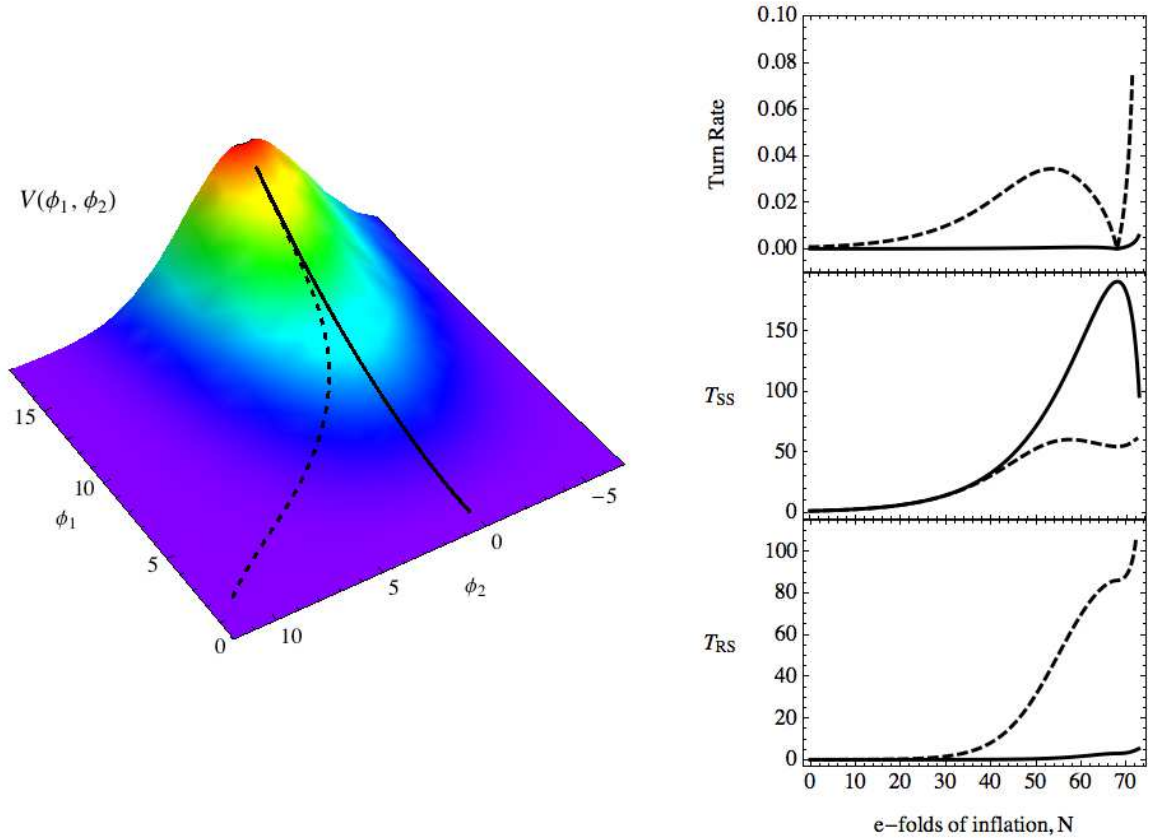}
\label{FigA}
\caption{For $|f_{NL}|$ to be large, the amount of sourcing of curvature modes by isocurvature modes ($T_{\mathcal{RS}}$) must be extremely sensitive to changes in the initial conditions and usually the amount of sourcing must also be moderate.  Above is an example of a trajectory (solid lines) that meets these two criteria. We use the potential $V(\phi_1, \phi_2) = \frac{1}{2} e^{-\lambda \phi_2^2} m^2 \phi_1^2$, which Byrnes \textit{et. al.} thoroughly investigated in \cite{ByrnesEtAl-2008}.   We set $\lambda = 0.05$ and illustrate the results for two trajectories: (1) one that starts at $(\phi_1^*,\phi_2^*) = (17,10^{-4})$, follows along the ridge, and turns ever so slightly at the end of inflation  (solid lines), and (2) a neighboring trajectory that starts only $|\Delta \boldsymbol{\phi}_*| = 0.01$ away in field space, but that eventually rolls off the narrow ridge (dashed lines).   The plot on the left shows the inflationary potential as a function of the fields, along with the two inflaton trajectories. The plots on the right show the turn rate, the relative amplitude of isocurvature modes ($T_{\mathcal{SS}}$), and the total amount of sourcing ($T_{\mathcal{RS}}$) as a function of $N$.  Here, the approximately 50-fold difference in the total sourcing stems more from the difference in the turn rates for the two trajectories, which both possess large isocurvature modes during all of inflation. Interestingly, the trajectory that rolls along the ridge (solid lines) produces $|f_{NL}| \sim 10^2$, while the neighboring trajectory (dashed lines) corresponds to $|f_{NL}| \approx 1$ \cite{ByrnesEtAl-2008}, visually illustrating the role of fine-tuning in achieving large non-Gaussianity.}
\end{minipage}
\end{figure*} 

Although we have not shown it here and the calculation is more difficult, similar qualitative conditions hold for general multi-field inflation; the main difference is that for multi-field inflation, the sourcing term analogous to $T_{\mathcal{RS}}$ is a vector, rather than a scalar. This explains why it has been difficult to find multi-field models of inflation that produce large non-Gaussianity: it is not easy to find inflationary scenarios that are so dramatically sensitive to the initial conditions and that involve moderate sourcing.  The interesting corollary of this is that some degree of fine-tuning is needed.  Fine-tuning is needed both to produce a potential where the mode sourcing is so sensitive to the initial conditions \textit{and} to start in the very narrow subset of initial conditions where both of the above conditions are satisfied. 

In Figure 2, we provide an example of an inflationary scenario that meets the two criteria associated with the bound in equation (\ref{cond on grad TRS}).  In the example below, an unstable ridge in the potential provides the perfect conditions for making the total amount of sourcing ($T_{\mathcal{RS}}$) so sensitive to the initial conditions.  The trajectory of interest (solid lines) rolls along the ridge and turns ever so slightly at the end of inflation, resulting in moderate sourcing and hence $|f_{NL}| \sim 10^2$.  By comparison, a neighboring trajectory (dashed lines) rolls off the ridge, experiences extremely strong sourcing, and produces $|f_{NL}| \approx 1$, which is just below the detection threshold for CMB experiments.

\subsection{Calculation of $\mathbf{e}_\perp^* \cdot \boldsymbol{\nabla}_* T_{\mathcal{RS}}$}

\label{perp grad T_RS}

Now we proceed to take the calculation of $\mathbf{e}_\perp^* \cdot \boldsymbol{\nabla}_* T_{\mathcal{RS}}$ as far as possible.  The work we present in this section is most applicable to analytically solvable models.  We also note that in this section, we explicitly indicate when the transfer function $T_{\mathcal{SS}}$ and certain other quantities are to be evaluated.  In particular, we use the superscript $e$ to denote that a quantity is to be evaluated at the end of inflation.

Calculating the term $\mathbf{e}_\perp^* \cdot \boldsymbol{\nabla}_* T_{\mathcal{RS}}$ is more difficult  because $T_{\mathcal{RS}}$ is often not a conservative function and because the field values at the end of inflation depend on the field values at horizon exit.  To act the operator $\boldsymbol{\nabla}_*$ on the expression for $T_{\mathcal{RS}}$ in equation~(\ref{transfer functions}), we change variables and rewrite the equation as a line integral expression of the fields.  We can convert equation (\ref{transfer functions}) into a line integral by working in the SRST limit and by replacing the function $\alpha = 2 \frac{\eta_{\perp}}{v}$ with its SRST counterpart $-2M_{\parallel\perp}$, which yields
\begin{align}
\label{T_RS in SRST}
T_{\mathcal{RS}} & =  - 2 \int_{\boldsymbol{\phi}_*}^{\boldsymbol{\phi}_e} T_{\mathcal{SS}}(\boldsymbol{\phi}_*,\boldsymbol{\phi})\,  \frac{\mathbf{e}_\perp^T  \boldsymbol{M} \, d\boldsymbol{\phi} }{\sqrt{2\epsilon}}. 
 \end{align}
If the integrand of $T_{\mathcal{RS}}$ is the gradient of a function, then operating $\boldsymbol{\nabla}_*$ on $T_{\mathcal{RS}}$ simply returns the integrand evaluated both at horizon exit and at the end of inflation, with the latter being times a matrix representing the sensitivity of the final field values to the initial field values (by virtue of the Chain Rule of calculus).  However, in general, the integrand of $T_{\mathcal{RS}}$ will not be the gradient of a function.  To account for this, we introduce a model-dependent function $\gamma$  to represent how much the integrand of $T_{\mathcal{RS}}$ in equation (\ref{T_RS in SRST}) deviates from being the gradient of a function.  Now operating $\boldsymbol{\nabla}_*$ on $T_{\mathcal{RS}}$ and using the new function $\gamma$, we obtain
\begin{widetext}
\begin{align}
\label{gradient T_RS}
 \boldsymbol{\nabla}_* T_{\mathcal{RS}} =  \left[\frac{2 \boldsymbol{M} \, \mathbf{e}_\perp}{\sqrt{2\epsilon}}  + \gamma \mathbf{e}_\perp \right]_* - \boldsymbol{\mathcal{X}} \left[\frac{2 \boldsymbol{M} \, \mathbf{e}_\perp}{\sqrt{2\epsilon}}  + \gamma \mathbf{e}_\perp \right]_e T_{\mathcal{SS}}^e + T_{\mathcal{RS}} \, \boldsymbol{\nabla}_* \left( \ln T_{\mathcal{SS}}^e \right)_{\boldsymbol{\phi}_e = const},
 \end{align}
 \end{widetext}
where
\begin{align}
\label{X matrix}
\mathcal{X}^i_{\, \, j} \equiv \frac{\partial C}{\partial \phi_i^*}  \frac{d \phi_j^e}{dC}.
\end{align} 
The matrix $\boldsymbol{\mathcal{X}}$ arises in the above expression due to the Chain Rule of calculus.  It captures how a change in the initial conditions at horizon exit affects the final values of the fields at the end of inflation.  Since only changes in the initial field vector that are off the trajectory will affect the final field values, the variable $C$ parametrizes motion orthogonal to the given trajectory; in other words, $C$ is constant along every unique trajectory of motion. 
The last new quantity we introduce in equation (\ref{gradient T_RS}) is the term $\boldsymbol{\nabla}_* \left( \ln T_{\mathcal{SS}}^e \right)_{\boldsymbol{\phi}_e = const}$, which means to take the gradient of $\ln T_{\mathcal{SS}}^e$ while holding the amplitude of the isocurvature modes at the end of inflation constant.  This term can be thought of as some sort of measure of the sensitivity of $T_{\mathcal{SS}}^e$ to the initial conditions.  This term arises from the fact that acting $\boldsymbol{\nabla}_*$ on $T_{\mathcal{RS}}$ in equation (\ref{T_RS in SRST}) involves differentiating under the integral, which is necessary since $T_{\mathcal{SS}}(\boldsymbol{\phi}_*, \boldsymbol{\phi})$ depends on $\boldsymbol{\phi}_*$.   

Considering equation (\ref{gradient T_RS}), the $\mathbf{e}_\parallel^*$ component is a model-independent expression simply because it can be related to the time-derivative of $T_{\mathcal{RS}}$, which has a model-independent form.  In fact, we already included this term in the first term on the right-hand side of equation (\ref{fNL general expression}) for $f_{NL}$, and hence we do not need to consider it further.  Thus, we only need to consider the $\mathbf{e}_\perp^*$ component of $\boldsymbol{\nabla}_* T_{\mathcal{RS}}$.  We emphasize that the form we assume for this component is best applicable to analytically solvable models and models where the coupling term $|\frac{\partial_1 \partial_2 V}{\partial_1 V \partial_2 V}| \ll 1$ or is approximately constant, as will become clearer later.  For non-analytic models, $\mathbf{e}_{\perp}^* \cdot \boldsymbol{\nabla}_* T_{\mathcal{RS}}$ can be evaluated numerically either directly from the expression for $T_{\mathcal{RS}}$ or via an alternative expression that we present later in this section.   In the remainder of this section, we discuss the three terms $\boldsymbol{\mathcal{X}}$, $T_{\mathcal{SS}}^e$, and $\gamma$, which arise in the expression for $\mathbf{e}_{\perp}^* \cdot \boldsymbol{\nabla}_* T_{\mathcal{RS}}$.  We first show that $\mathcal{\mathcal{X}}$ has a model-independent form.  Then, we discuss the transfer function $T_{\mathcal{SS}}^e$, how it affects $T_{\mathcal{RS}}$ and the gradient of $T_{\mathcal{RS}}$, and how it determines the model-dependent term $\gamma$.  Along the way, we consider the exact solutions for product and sum potentials, casting previous results for $f_{NL}$ for these models \cite{VernizziAndWands-2006,ChoiEtAl-2007} in a more geometrically and physically transparent form.  
 
Now we find the matrix $\boldsymbol{\mathcal{X}}$ and prove that it has a model-independent form.  First, since $C$ is constant along a given trajectory, 
\begin{align}
C' = \boldsymbol{\phi}'  \cdot \boldsymbol{\nabla} C = 0. 
\end{align}
Therefore, $\boldsymbol{\nabla} C$ must be proportional to $\mathbf{e}_{\perp}$ --- that is,
\begin{align}
\label{gradient C}
\boldsymbol{\nabla} C = |\boldsymbol{\nabla} C| \, \mathbf{e}_\perp.
\end{align}
Next, consider $\frac{d \boldsymbol{\phi}_e}{dC}$.  Since $C$ is constant along a trajectory, a change in $C$ corresponds to motion orthogonal to the trajectory, and hence $\frac{d \boldsymbol{\phi}^e}{dC}$ must be parallel to $\mathbf{e}_\perp^e$.  Now combining this fact with equation (\ref{gradient C}) and with
\begin{align}
1 = \frac{dC}{dC} = \frac{d \boldsymbol{\phi}^e}{dC} \cdot \boldsymbol{\nabla}_e C
\end{align}
implies that $\frac{d \boldsymbol{\phi}^e}{dC} =  |\boldsymbol{\nabla} C| _e^{-1} \mathbf{e}_\perp^e$.  Substituting this result and equation (\ref{gradient C}) into equation (\ref{X matrix}) yields
\begin{align}
\label{X matrix in terms of zeta v1}
\boldsymbol{\mathcal{X}} = \frac{ |\boldsymbol{\nabla} C| _*} {|\boldsymbol{\nabla} C| _e} \mathbf{e}_\perp^* (\mathbf{e}_{\perp}^e)^T.
\end{align}
Finally, we express the ratio of the norms of the gradients of $C$ at horizon exit and the end of inflation in terms of a physical quantity: the relative amplitude of the entropy modes.  Since $\delta C = \boldsymbol{\delta\phi} \cdot \boldsymbol{\nabla} C$, then for a given variation in the trajectory, $\delta C$, we have
\begin{align}
\label{relation btw zeta and entropy modes}
 |\boldsymbol{\nabla} C| _* \, \delta \phi_{\perp}^* = |\boldsymbol{\nabla} C| _e \, \delta \phi_{\perp}^e.
\end{align}
Combining equations (\ref{X matrix in terms of zeta v1}) and (\ref{relation btw zeta and entropy modes}), we finally arrive at the model-independent expression
\begin{align}
\label{X matrix in terms of zeta}
\boldsymbol{\mathcal{X}} & = \left(\frac{\delta \phi_{\perp}^e}{\delta \phi_{\perp}^* }\right)  \mathbf{e}_\perp^* (\mathbf{e}_{\perp}^e)^T  =\sqrt{\frac{2\epsilon_e}{2\epsilon_*} \,}T_{\mathcal{SS}}^e \,  \mathbf{e}_\perp^*  (\mathbf{e}_{\perp}^e)^T.
\end{align}

This interesting result shows that the sensitivity of the final field values to the initial field values can be given very simply in terms of the relative growth or decay of entropy modes.  In other words, the evolution of entropy modes mirrors whether neighboring trajectories converge or diverge over time.  In scenarios where neighboring trajectories converge (``attractor solutions''), the entropy modes decay.  However, when neighboring trajectories diverge, the entropy modes grow.   That such a relationship should hold between the convergence/divergence of neighboring trajectories and the evolution of entropy modes makes sense.  From a geometrical perspective, we intuitively expect that a positive curvature along the entropic direction focuses neighboring trajectories, whereas a negative curvature creates a hill or ridge in the potential, causing neighboring trajectories to diverge.  But we also know that the curvature along the entropic direction determines the evolution of entropy modes.  By equation (\ref{mode EoMs}), the entropy modes grow when $M_{\perp \perp} < 0$ and decay when $M_{\perp \perp}>0$, and how quickly they do so depends on the magnitude of the curvature.   Combining these two facts together, we could have concluded that the divergence/convergence of neighboring trajectories must correlate with the growth/decay of entropy modes, without even deriving this result.  Nonetheless, equation (\ref{X matrix in terms of zeta}) gives the precise relationship explicitly.

Now substituting equation (\ref{X matrix in terms of zeta}) into equation (\ref{gradient T_RS}) and projecting the result onto $\mathbf{e}_\perp^*$, we obtain 
\begin{widetext}
\begin{align}
\label{grad T_RS final}
\sqrt{2\epsilon_*} \, \mathbf{e}_{\perp}^* \cdot  \boldsymbol{\nabla}_* T_{\mathcal{RS}} = \left(2 M^*_{\perp \perp} +   \sqrt{2\epsilon_*}  \, \gamma_* \right) -  \left(2 M^e_{\perp \perp} + \sqrt{2\epsilon_e}  \, \gamma_e  \right) (T_{\mathcal{SS}}^e)^2 + \sqrt{2\epsilon_*} \, T_{\mathcal{RS}} \, \mathbf{e}_{\perp}^* \cdot  \boldsymbol{\nabla}_*  \left( \ln T_{\mathcal{SS}}^e \right)_{\boldsymbol{\phi}_e = const}.
\end{align}
\end{widetext}
The above equation shows that the sensitivity of $T_{\mathcal{RS}}$ to the initial conditions is determined by $M_{\perp \perp}$, $\epsilon$, $T_{\mathcal{SS}}^e$, $T_{\mathcal{RS}}$, and the model-dependent factor $\gamma$.  The above relation gives us a nice way to understand when the sourcing is very sensitive to the initial conditions, based on the geometrical and physical attributes of an inflationary model.

Finally, we consider the model-dependent quantity $\gamma$.  Recall that we defined $\gamma$ so that it is zero whenever the integrand of $T_{\mathcal{RS}}$ is the gradient of a function.  This occurs for product potentials, defined as
\begin{align}
V = V_1(\phi_1) V_2(\phi_2),
\end{align}
and can be attributed to the fact that in these models, the two fields evolve independently of each other.  $\gamma$ is therefore non-zero whenever the evolutions of the two fields influence each other.

We can see that $\gamma$ is zero for product potentials as follows. For product potentials, the isocurvature mass equals 
\begin{align}
\label{M perp perp for prod potls}
\beta = M_{\parallel \parallel} - M_{\perp \perp} = (\tan \theta - \cot \theta) M_{\parallel \perp}, 
\end{align}
where $\theta$ is the polar coordinate in the $(\phi_1',\phi_2')$ plane, 
\begin{align}
\tan \theta \equiv \frac{\phi_2'}{\phi_1'}.
\end{align}
Using $-M_{\parallel \perp} \approx \frac{\eta_\perp}{v} =  \theta'$ and plugging equation (\ref{M perp perp for prod potls}) into equation (\ref{transfer functions}), one finds that the transfer function $T_{\mathcal{SS}}^e$ for these models can be approximated by
\begin{align}
\label{T_SS for mult potls}
T_{\mathcal{SS}}^e = \frac{\sin \theta_e \cos \theta_e}{\sin \theta_* \cos \theta_*}.
\end{align}
Substituting $\frac{\eta_\perp}{v} = \theta '$ and equation (\ref{T_SS for mult potls}) into equation (\ref{transfer functions}) yields
\begin{align}
T_{\mathcal{RS}} & = \int_{N_*}^{N_e} 2 \theta' \, \frac{\sin \theta \cos \theta}{\sin \theta_* \cos \theta_*} \, dN, \nonumber\\
 & = \frac{1}{\sin \theta_* \cos \theta_*} \int_{N_*}^{N_e} \frac{d}{dN} \left(\sin^2 \theta\right) dN,
\end{align}
which integrates to give
\begin{align}
T_{\mathcal{RS}} & = - \tan \theta_* + \tan \theta_e T_{\mathcal{SS}}^e \nonumber \\
& = \frac{1}{\sin \theta_* \cos \theta_*} \left(\sin^2 \theta_e - \sin^2 \theta_*\right).
\end{align}
Now we take the gradient of the above transfer function and use that
\begin{align}
\label{grad of unit vec}
\boldsymbol{\nabla}  \mathbf{e}_\parallel = \boldsymbol{\nabla} (\cos \theta, \sin \theta) = - \frac{\boldsymbol{M} \mathbf{e}_{\perp}}{\sqrt{2\epsilon}} \mathbf{e}_{\perp}^T
\end{align}
in the SRST limit for any two-field model of inflation.  Finally, projecting the result onto $\mathbf{e}_{\perp}^*$ and using equation (\ref{M perp perp for prod potls}), we find
\begin{widetext}
\begin{align}
\label{grad T_RS final for mult potls}
\sqrt{2\epsilon_*} \, \mathbf{e}_{\perp}^* \cdot  \boldsymbol{\nabla}_* T_{\mathcal{RS}} =  & \, 2 M^*_{\perp \perp} -  2 M^e_{\perp \perp}  (T_{\mathcal{SS}}^e)^2 +  \sqrt{2\epsilon_*} \, T_{\mathcal{RS}} \, \mathbf{e}_{\perp}^* \cdot  \boldsymbol{\nabla}_*  \left( \ln T_{\mathcal{SS}}^e \right)_{\boldsymbol{\phi}_e = const}, \nonumber \\
 = &  \left[2 + (\cot \theta_* - \tan \theta_*)T_{\mathcal{RS}}\right] M^*_{\perp \perp} -  2 M^e_{\perp \perp}  (T_{\mathcal{SS}}^e)^2, \nonumber \\
 = &  \left[2 + \left(\frac{M_{\perp \perp}^* - M_{\parallel \parallel}^*}{M_{\parallel \perp}^*} \right) T_{\mathcal{RS}}\right] M^*_{\perp \perp} -  2 M^e_{\perp \perp}  (T_{\mathcal{SS}}^e)^2. 
 \end{align}
\end{widetext}
where again $T_{\mathcal{SS}}^e$ for product potentials is given by equation (\ref{T_SS for mult potls}).  Comparing equation (\ref{grad T_RS final for mult potls}) to equation (\ref{gradient T_RS}) indeed shows that $\gamma = 0$.  

%From equation (\ref{grad T_RS final for mult potls}), we can see that $\sqrt{2\epsilon_*} \, \mathbf{e}_{\perp}^* \cdot  \boldsymbol{\nabla}_* T_{\mathcal{RS}} $ will be large in magnitude for product potentials when the isocurvature mass is large relative to the turn rate at horizon exit (or equivalently, either one of the two fields starts with far more kinetic energy than the other), $M_{\perp \perp}^*$ is large (which is usually not permitted in standard slow-roll inflation), and/or when $M_{\perp \perp}^e (T_{\mathcal{SS}}^e)^2$ is large. If we assume that $M_{\perp \perp}^e$ is at most of order O(1)-O(10), then for the last condition to be satisfied requires that the amplitude of isocurvature modes at the end of inflation be large.  But by equation (\ref{T_SS for mult potls}), $T_{\mathcal{SS}}^e \ge 1$ only if the asymmetry in how much kinetic energy the two fields have either diminishes with time or at least holds steady.  

Now for the general case of two-field inflation, we can use a similar procedure to find the model-dependent term $\gamma$.  As the transfer function $T_{\mathcal{SS}}$ determines how much the integrand of $T_{\mathcal{RS}}$ deviates from being the gradient of a function and hence determines $\gamma$, we start by considering $T_{\mathcal{SS}}$.  We begin by finding a general expression for the transfer function $T_{\mathcal{SS}}$.  Starting from equation (42) in \cite{MukhanovAndSteinhardt-1997}, after some algebra, we can show that this implies that $T_{\mathcal{SS}}^e$ takes the form
\begin{align}
T_{\mathcal{SS}}^e = & \left(\frac{\sin \theta_e \cos \theta_e}{\sin \theta_* \cos \theta_*}\right)  \, \exp \left[\int_{N_*}^{N_e} \frac{M_{12}}{\sin \theta \cos \theta} dN \right], \nonumber \\
= & \left(\frac{\sin \theta_e \cos \theta_e V_e}{\sin \theta_* \cos \theta_* V_*}\right) s(\boldsymbol{\phi}_*,\boldsymbol{\phi_e})
\end{align} 
where $M_{12} \equiv \partial_1 \partial_2 \ln V$ and
\begin{align}
\label{s integral}
s(\boldsymbol{\phi}_*,\boldsymbol{\phi_e}) \equiv \exp \left[- \int_{\boldsymbol{\phi}_*}^{\boldsymbol{\phi}_e} \left(\frac{V \partial_1 \partial_2 V}{\partial_1 V \partial_2 V}\right) \boldsymbol{\nabla} \ln V \cdot d\boldsymbol{\phi} \right].
\end{align}
This means that whenever $\left(\frac{V \partial_1 \partial_2 V}{\partial_1 V \partial_2 V}\right)$ is a constant, $s(\boldsymbol{\phi}_*,\boldsymbol{\phi_e})$ becomes analytic and hence $T_{\mathcal{SS}}$ becomes analytic.   For product potentials, $M_{12} = 0$, reproducing the result we derived in equation (\ref{T_SS for mult potls}).  For sum potentials, defined as
\begin{align}
V = V_1(\phi_1) + V_2(\phi_2),
\end{align} 
the coupling term $\partial_1 \partial_2 V = 0$, and so
\begin{align}
\label{T_SS for sum potls}
T_{\mathcal{SS}}^e = \frac{\sin \theta_e \cos \theta_e V_e}{\sin \theta_* \cos \theta_* V_*}.
\end{align}
Equation (\ref{T_SS for sum potls}) for $T_{\mathcal{SS}}^e$ can also be used as an approximation for scenarios in which $\left|\frac{V \partial_1 \partial_2 V}{\partial_1 V \partial_2 V}\right| \ll 1$ during all of inflation.  In the more general case where $\left|\frac{V \partial_1 \partial_2 V}{\partial_1 V \partial_2 V}\right|$ is approximately constant during inflation, $T_{\mathcal{SS}}^e$ can be approximated by an analytic function similar to equation (\ref{T_SS for sum potls}), but possessing additional powers of $(V_e/V_*)$.  And whenever $\left(\frac{V \partial_1 \partial_2 V}{\partial_1 V \partial_2 V}\right)$ is approximately constant --- which is automatically true for all product and sum potentials --- the following relation holds: 
\begin{align}
\sqrt{2\epsilon_*} \, \mathbf{e}_{\perp}^* \cdot & \boldsymbol{\nabla}_* (\ln T_{\mathcal{SS}}^e)_{\boldsymbol{\phi}_e = const} =  (\cot \theta_* - \tan \theta_*) \, M^*_{\perp \perp},
%& =  \left[\frac{M_{\perp \perp}^* - M_{\parallel \parallel}^* + \left(\left(\frac{V \partial_1 \partial_2 V}{\partial_1 V \partial_2 V}\right)_* - 1\right) 2\epsilon_*}{M_{\parallel \perp}^*}\right] M_{\perp \perp}^*,
\end{align}
as gradients of functions of $V$ do not contribute to $\mathbf{e}_{\perp}^* \cdot \boldsymbol{\nabla}_* (\ln T_{\mathcal{SS}}^e)_{\boldsymbol{\phi}_e = const}$.  Otherwise, when the term $\left(\frac{V \partial_1 \partial_2 V}{\partial_1 V \partial_2 V}\right)$ is not a constant, this adds extra terms to  $\sqrt{2\epsilon_*} \, \mathbf{e}_{\perp}^* \cdot \boldsymbol{\nabla}_* (\ln T_{\mathcal{SS}}^e)_{\boldsymbol{\phi}_e = const}$ that contribute to $f_{NL}$. 

Now plugging the general expression for $T_{\mathcal{SS}}$ into equation (\ref{transfer functions}) and integrating by parts again using the fact that $(\sin^2 \theta)' = 2 \sin \theta \cos \theta \theta'$, we obtain
\begin{align}
\label{T_RS int by parts}
T_{\mathcal{RS}} = & - \tan \theta_* + \tan \theta_e T_{\mathcal{SS}}^e - \\
&  \frac{1}{\sin \theta_* \cos \theta_* V_*} \int_{N_*}^{N_e} \sin^2 \theta \frac{d}{dN}\left( V s(N_*,N)\right) dN.  \nonumber
\end{align}
Now the perpendicular component of the gradient of $T_{\mathcal{RS}}$ can be calculated directly using the above equation,\footnote{From equation (\ref{T_RS int by parts}), we can also derive an upper limit for $T_{\mathcal{RS}}$ for general two-field inflation whenever the SRST limit is a valid approximation.  Using the fact that $\sin^2 \theta \le 1$, we find that from equation (\ref{T_RS int by parts}), \begin{align} T_{\mathcal{RS}} \le \cot \theta_* - \cot \theta_e T_{\mathcal{SS}}^e.\end{align}} 
 where equation (\ref{grad of unit vec}) comes in handy.  In particular, for sum potentials, the integral in equation (\ref{T_RS int by parts}) evaluates to $V_2^e - V_2^*$, yielding
\begin{align}
\label{T_RS int by parts for sum potls}
T_{\mathcal{RS}} = & - \tan \theta_* + \frac{V_2^*}{\sin \theta_* \cos \theta_* V_*} \nonumber \\ & + \left(\tan \theta_e - \frac{V^e_2}{\sin \theta_e \cos \theta_e V_e}\right) T_{\mathcal{SS}}^e. 
\end{align}
For all other potentials,  the integral in equation (\ref{T_RS int by parts}) typically must be computed numerically, which is still usually the fastest route to finding $\mathbf{e}_{\perp}^* \cdot \boldsymbol{\nabla}_* T_{\mathcal{RS}}$.    However, we instead use the above expression to find the model-dependent term $\gamma$.  Taking the gradient of equation (\ref{T_RS int by parts}) and factoring it into the form of equation (\ref{gradient T_RS}), we conclude that $\gamma$ can be written as
\begin{widetext}
\begin{align}
\label{gamma result}
%\gamma_e = \left[- \tan \theta_* \, \mathbf{e}_{\perp}^* \cdot \boldsymbol{\nabla}_* \ln (s(\boldsymbol{\phi}_*,\boldsymbol{\phi}_e))  - \frac{1}{\sin \theta_* \cos \theta_*V_*}  \mathbf{e}_{\perp}^* \cdot \boldsymbol{\nabla}_* \left(\left(1 - \ln (s(\boldsymbol{\phi}_*,\boldsymbol{\phi}_e))  \right) \int_{\boldsymbol{\phi}_*}^{\boldsymbol{\phi}_e} \sin^2 \theta \, \boldsymbol{\nabla} \left(V \, s(\boldsymbol{\phi}_*,\boldsymbol{\phi}_e)\right) \cdot d\boldsymbol{\phi} \right)\right]_{\boldsymbol{\phi}_e = constant}, \\
\gamma_e = \left[-\tan \theta_e \, \mathbf{e}_{\perp}^e \cdot \boldsymbol{\nabla}_e \ln (s(\boldsymbol{\phi}_*,\boldsymbol{\phi}_e))  + \frac{1}{\sin \theta_* \cos \theta_*V_*}  \mathbf{e}_{\perp}^e \cdot \boldsymbol{\nabla}_e \left(\int_{\boldsymbol{\phi}_*}^{\boldsymbol{\phi}_e} \sin^2 \theta \, \boldsymbol{\nabla} \left(V \, s(\boldsymbol{\phi}_*,\boldsymbol{\phi})\right) \cdot d\boldsymbol{\phi} \right)\right]_{\boldsymbol{\phi}_* = constant},
\end{align}
\end{widetext}
where the gradients are taken with respect to holding the initial field vector constant.  The resulting expression for $\gamma$ at horizon exit has the same functional form, of course.

For sum potentials, the function $s(\boldsymbol{\phi}_*,\boldsymbol{\phi})$ vanishes, leaving only the perpendicular component of the gradient of the integral $\int_{\boldsymbol{\phi}_*}^{\boldsymbol{\phi}_e} \sin^2 \theta \, \boldsymbol{\nabla} \, V \cdot d\boldsymbol{\phi} = V_2^e - V_2^*$.  Hence we find that $\gamma = -\sqrt{2\epsilon}$.    Therefore, the expression for $\mathbf{e}_\perp^* \cdot \boldsymbol{\nabla}_* T_{\mathcal{RS}}$ for sum potentials can be written as
\begin{widetext}
\begin{align}
\label{grad T_RS final for add potls}
\sqrt{2\epsilon_*} \, \mathbf{e}_{\perp}^* \cdot  \boldsymbol{\nabla}_* T_{\mathcal{RS}} = & (2 M^*_{\perp \perp} - 2\epsilon_*) -  (2 M^e_{\perp \perp} - 2\epsilon_e) (T_{\mathcal{SS}}^e)^2 +  \sqrt{2\epsilon_*} \, T_{\mathcal{RS}} \, \mathbf{e}_{\perp}^* \cdot  \boldsymbol{\nabla}_*  \left( \ln T_{\mathcal{SS}}^e \right)_{\boldsymbol{\phi}_e = const}, \nonumber \\
 = &  \left[2 + (\cot \theta_* - \tan \theta_*)T_{\mathcal{RS}}\right] M^*_{\perp \perp} - 2\epsilon_*-  (2 M^e_{\perp \perp} - 2\epsilon_e) (T_{\mathcal{SS}}^e)^2, \nonumber \\
 = &  \left[2 + \left(\frac{M_{\perp \perp}^* - M_{\parallel \parallel}^* - 2\epsilon_*}{M_{\parallel \perp}^*}\right) T_{\mathcal{RS}}\right] M^*_{\perp \perp} - 2\epsilon_*-  (2 M^e_{\perp \perp} - 2\epsilon_e) (T_{\mathcal{SS}}^e)^2,  
\end{align}
\end{widetext}
where $T_{\mathcal{SS}}^e$ is given by equation (\ref{T_SS for sum potls}).
%The conditions for large $|\sqrt{2\epsilon_*} \, \mathbf{e}_{\perp}^* \cdot  \boldsymbol{\nabla}_* T_{\mathcal{RS}} |$ for sum potentials are thus physically similar to those for product potentials, with the exception that the coefficient in front of $(T_{\mathcal{SS}}^e)^2$ is guaranteed to possess at least one term of order unity, if we assume that the end of inflation corresponds to $\epsilon_e = 1$.  We further discuss the complete set of conditions for large $|f_{NL}|$ for both product and sum potentials in the next section, connecting work initially done by Byrnes \textit{et. al.} \cite{ByrnesEtAl-2008} more closely to the geometrical and physical features such models need in order to produce large $|f_{NL}|$.

For all other potentials, equations (\ref{grad T_RS final}),  (\ref{s integral}) , and (\ref{gamma result}) represent the prescription for calculating $\mathbf{e}_{\perp}^* \cdot \boldsymbol{\nabla}_* T_{\mathcal{RS}}$.  For weak coupling among the fields, we expect $\gamma$ to be of order the slow-roll parameters.  However, this term may be larger in the limit of strong coupling.

\subsection{Conditions for Large $f_{NL}$}

As we showed in Section (\ref{bispec initial calc}), if the power spectra are nearly scale-invariant, the magnitude of $f_{NL}$ can be greater than unity only if $|\sin \Delta_N^2 \cos \Delta_N^2 \sqrt{2\epsilon_*} \,  \mathbf{e}_{\perp}^* \cdot \boldsymbol{\nabla}_* T_{\mathcal{RS}}| \gtrsim 1$.  Satisfying this bound requires that two conditions be met: (1) that $T_{\mathcal{RS}}$ be extremely sensitive to changes in the initial conditions perpendicular to the given trajectory and (2) that the amount of sourcing be non-zero.   Typically, it also means that the sourcing of curvature modes by isocurvature modes must be moderate, although this is not strictly required; rather, very weak sourcing ($\sin \Delta_N \ll 1$) or very strong sourcing ($\cos \Delta_N \gg 1$) simply makes the bound in equation (\ref{cond on grad TRS}) extremely hard to satisfy.  

The first condition can be understood very simply: in order to produce a large degree of skew in the primordial fluctuations, perturbations off the classical trajectory must move the inflaton onto neighboring trajectories that experience very different dynamics for the field perturbations. To satisfy this condition requires that the profiles of the turn rate and/or of the relative amplitude of isocurvature modes ($T_{\mathcal{SS}}$) be dramatically different for neighboring trajectories.  The second condition, that the sourcing must be non-zero, requires that the turn rate not be zero for all of inflation.  And to achieve moderate sourcing, which is usually needed to satisfy the bound in equation (\ref{cond on grad TRS}), the larger $T_{\mathcal{SS}}$ is, the smaller the turn rate must be, and vice versa.  

One question that naturally arises is whether combining the need for moderate sourcing with the requirement that the total sourcing be dramatically different among neighboring trajectories gives us any constraints on the possible ways that the turn rate and/or $T_{\mathcal{SS}}$ can vary and still produce large $|f_{NL}|$.  The answer to this question is yes.  We can best see this by considering the sourcing function $T_{\mathcal{RS}}$.   Consider the case where the amplitude of isocurvature modes never exceeds its value at horizon exit, \ie, $T_{\mathcal{SS}} (N_*, N) \le 1$.  In this case, $T_{\mathcal{RS}}$ can never exceed
\begin{align}
\label{TRS bound}
T_{\mathcal{RS}} \le 2 \int_{N_*}^{N_e} \theta' \, dN = 2 (\theta_e - \theta_*),
\end{align}
where again $\theta$ is the polar angle for the field velocity vector.
For the most common scenarios, the field velocity vector does not turn through an angle of more than $90^o$, yielding a bound of $T_{\mathcal{RS}} \le \pi$. Let us compare this bound of $T_{\mathcal{RS}} \le \pi$ with a numerical example.  If $\epsilon_* = 0.02$ and we assume nearly scale-invariant scalar and tensor spectra, then we need $\mathbf{e}_{\perp}^* \cdot \boldsymbol{\nabla}_* T_{\mathcal{RS}} \gtrsim 60$ in order to produce $|f_{NL}| \approx 3$.  This seems extremely difficult to achieve given the bound of $T_{\mathcal{RS}} \le \pi$ \textit{and} the need for moderate sourcing.  Now if within this set of scenarios, we consider the subset where $M_{\perp \perp} \ge 0$ during all of inflation, then neighboring trajectories must converge (or at least not diverge) over time.  Since neighboring trajectories remain close to each other during all of inflation, $T_{\mathcal{RS}}$ cannot differ widely among neighboring trajectories without discontinuous or other extreme features in the potential that violate the SRST conditions. To be more precise, this would require the speed-up rate, turn rate, and/or entropy mass to be hugely varying in the direction orthogonal to the given trajectory, which effectively constitutes a violation of higher-order SRST parameters. And on an intuitive level, it is not possible for a neighboring trajectory to have a much larger or smaller turn rate for a substantial period of time without having the two trajectories diverge,\footnote{We make the usual unstated assumption that there are no classical degeneracies in the gradient of $\ln V$, which means that trajectories cannot cross each other.}  which violates $M_{\perp \perp} \ge 0$.  Therefore, since large non-Gaussianity cannot be produced under these conditions, either $T_{\mathcal{SS}} > 1$ and/or $M_{\perp \perp} < 0$ at least sometime during inflation.  

In Section (\ref{perp grad T_RS}), we took the calculation of $\sqrt{2\epsilon_*} \, \mathbf{e}_{\perp}^* \cdot \boldsymbol{\nabla}_* T_{\mathcal{RS}}$ as far as possible, trying to better understand when this term is large in magnitude in analytically solvable and similar models.  From equation (\ref{grad T_RS final}), we found that $\sqrt{2\epsilon_*} \, \mathbf{e}_{\perp}^* \cdot  \boldsymbol{\nabla}_* T_{\mathcal{RS}} $ will be large in magnitude if at least one of the following three conditions is met:
\begin{enumerate}
\item  $2M_{\perp \perp}^* - \sqrt{2\epsilon_*}\gamma_*$ is large in magnitude,
\item  $(2M_{\perp \perp}^e - \sqrt{2\epsilon_e}\gamma_e) (T_{\mathcal{SS}}^e)^2$ is large in magnitude, or
\item $T_{\mathcal{SS}}^e$ is very sensitive to changes in the initial conditions orthogonal to the inflaton trajectory.
\end{enumerate}
In conventional slow-roll, the magnitudes of $M_{\perp \perp}^*$ and the other slow-roll parameters at horizon exit are significantly less than unity, so only the latter two conditions can be satisfied.  If we assume conventional slow-roll at horizon exit and additionally that the magnitudes of  $M_{\perp \perp}^e$ and $\gamma_e$ are at most of order \textit{O}(1)-\textit{O}(10), then the second condition above requires that $T_{\mathcal{SS}}^e$ is at least of order unity.  If we tighten the constraints even further, requiring that $|M_{\perp \perp}| \ll 1$ during all of inflation, then the second condition becomes even more stringent, requiring that $T_{\mathcal{SS}}^e$ be very large, at least of order $O$($100$).  

Interestingly, we can show that essentially the same conditions for large  $\sqrt{2\epsilon_*} \mathbf{e}_{\perp}^* \cdot \boldsymbol{\nabla}_* T_{\mathcal{RS}}$ arise via an alternative approach.  Anytime $\boldsymbol{\nabla} N|_{\boldsymbol{\phi}_e = constant} = \mathbf{F}$ --- which\footnote{By this expression, we mean that the gradient of $N$ evaluated when holding the final field vector constant equals $\mathbf{F}$.} includes all scenarios in which $N$ is a function of only the initial and final fields --- we have
\begin{align}
\boldsymbol{\nabla}_* N = - \mathbf{F}_* + \boldsymbol{\mathcal{X}} \mathbf{F}_e,
\end{align}
and hence by equations (\ref{nabla N formula}) and (\ref{X matrix in terms of zeta}),
\begin{align}
T_{\mathcal{RS}} =  - \left[\sqrt{2\epsilon} F_\perp \right]_* + \left[ \sqrt{2\epsilon} F_{\perp}\right]_e T_{\mathcal{SS}}^e.
\end{align}
Now taking the perpendicular component of the gradient of $T_{\mathcal{RS}}$, we obtain
\begin{widetext}
\begin{align}
\label{T_RS for general delta N}
\sqrt{2\epsilon_*} \, \nabla_\perp^* T_{\mathcal{RS}} =   -\left[\sqrt{2\epsilon} \nabla_\perp \left(\sqrt{2\epsilon} F_\perp \right)\right]_* + \left[\sqrt{2\epsilon} \nabla_\perp \left(\sqrt{2\epsilon} F_\perp \right)\right]_e (T_{\mathcal{SS}}^e)^2 + \left[ \sqrt{2\epsilon} F_\perp \right]_e \sqrt{2\epsilon_*} \nabla_\perp^* T_{\mathcal{SS}}^e,
\end{align}
\end{widetext}
where $\nabla_{\perp} \equiv \mathbf{e}_{\perp} \cdot \boldsymbol{\nabla}$.  So under similar assumptions, it again appears that the amplitude of isocurvature modes at the end of inflation and the sensitively of $T_{\mathcal{SS}}^e$ to a change in initial conditions orthogonal to the given classical trajectory largely control the magnitude of $f_{NL}$.

Let us now consider the above conditions in the context of the two most popular categories of models: product potentials and sum potentials.  Expressions for $f_{NL}$ for product potentials and sum potentials were first found by \cite{ChoiEtAl-2007} and \cite{VernizziAndWands-2006}, respectively.  Our versions of the same expressions --- equation (\ref{fNL general expression}) coupled either equation (\ref{grad T_RS final for mult potls}) or (\ref{grad T_RS final for add potls}), respectively --- give somewhat more direct insight into how $f_{NL}$ depends on the physical and geometrical features of an inflationary model.   
The conditions for large bispectra in both product and sum potentials were discovered by Byrnes \textit{et. al.} \cite{ByrnesEtAl-2008}.  For product potentials, they found that $f_{NL}$ will be large in magnitude when either one of the two fields starts with far more kinetic energy than the other (either $\cot \theta_* \gg 1$ or $\tan \theta_* \gg 1$) and when the asymmetry in the kinetic energies of the two fields diminishes significantly by the end of inflation.   The reason that these two conditions produce large bispectra in product potentials is that they together guarantee three things: that $T_{\mathcal{SS}}^e \gg 1$ (which follows from equation (\ref{T_SS for mult potls})), that $T_{\mathcal{SS}}^e$ is very sensitive to changes in the initial conditions orthogonal to the inflaton trajectory, and that the turn rate is small yet significant enough to produce moderate sourcing.  
Interestingly, this means that for product potentials, both the second and third conditions for large $\sqrt{2\epsilon_*} \, \mathbf{e}_{\perp}^* \cdot  \boldsymbol{\nabla}_* T_{\mathcal{RS}} $ are always simultaneously satisfied, as the third condition combined with the requirement that the total sourcing is moderate guarantees the second condition, and vice versa.  For sum potentials, the conditions for large bispectra are a bit more complicated to untangle but end up being similar; however, the expression for $T_{\mathcal{SS}}^e$ contains a factor of $\frac{V_e}{V_*}$.   Although we state the conditions for large $|f_{NL}|$ differently and slightly extend them in scope by relaxing constraints on $M_{\perp \perp}$, these conditions otherwise agree with those uncovered by Byrnes \textit{et. al.} \cite{ByrnesEtAl-2008}.

As we concluded in Section (\ref{bispec initial calc}) that large non-Gaussianity requires the profiles of the turn rate and/or $T_{\mathcal{SS}}$ to be dramatically different for neighboring trajectories, we might worry why we did not uncover any explicit conditions that involve the turn rate.  There are a few reasons for this.  First, the lowest-order time variation of the turn rate is a model-dependent function of the isocurvature mass, $\beta = M_{\parallel \parallel} - M_{\perp \perp}$ and of the coupling between the fields, so the turn rate is implicitly included in the above conditions.  Second, this should not worry us as the turn rate is constrained not to be too large by the constraints on scale-invariance and is constrained not to be too small by the need for moderate sourcing.  And third, the difference in the turn rate between two neighboring trajectories cannot be large for a sustained amount of time without neighboring trajectories diverging, so $M_{\perp \perp}$ itself also tells us whether larger variations in the turn rate between neighboring trajectories are possible. %Therefore, if the turn rate is smaller than of order unity for neighboring trajectories (which we assumed by invoking the slow-turn approximation), then this means that the isocurvature modes must be significantly large during at least some part of inflation in order to both satisfy the requirement for moderate sourcing and to guarantee that $\sqrt{2\epsilon_*} \mathbf{e}_{\perp}^* \cdot \boldsymbol{\nabla}_* T_{\mathcal{RS}}$ is large.  

Now we consider the geometric implications for the inflationary potential.   We argued earlier that it is not possible to achieve large non-Gaussianity during inflation if both $T_{\mathcal{SS}}(N_*,N) \le 1$ and $M_{\perp \perp} \ge 0$ during all of inflation.  Therefore, one or both of the conditions must be violated sometime during inflation to produce large non-Gaussianity.  If we make the conventional assumptions that $\epsilon_* \ll 1$, $\epsilon$ increases significantly (but not necessarily monotically) in order to end inflation, and that $\epsilon$ never drops below its value at horizon exit, then the only way to satisfy $T_{\mathcal{SS}}(N_*,N) \ge 1$ is for the entropy modes to grow at some point during inflation.\footnote{If we relax the assumption that $\epsilon$ never decreases below its value at horizon exit, then it may be possible to produce large $|f_{NL}|$ in two-field inflation without any negative curvature along the entropic direction.  Indeed, Byrnes \textit{et. al.} \cite{ByrnesEtAl-2008b} showed that this is possible for a three-field model: a two-component hybrid inflation model, where we are counting the waterfall field as the third field.  During the first phase of inflation, which is governed by a vacuum-dominated sum potential and during which only two of the fields are active, $\epsilon$ exponentially decays after modes exit the horizon.  The exponential decay of $\epsilon$ causes the isocurvature modes to grow dramatically during the initial phase of inflation, which produces large non-Gaussianity even before the waterfall field comes into play.  This large exponential decay in $\epsilon$ then circumvents the absolute need for a negative entropy mass at some point during inflation.  However, for this model to then to be viable, a third field, the waterfall field with its associated negative mass, is needed to end inflation.  We also note that double inflation models with very high mass ratios do violate the assumption that $\epsilon$ never drops below $\epsilon_*$ and hence exceed the bound $T_{\mathcal{RS}} \le \pi$, but nonetheless, they do not to produce large non-Gaussianity.}   As we showed earlier, the entropic modes grow when the curvature along the entropic direction is negative, or equivalently, when neighboring trajectories diverge. Hence under conventional assumptions, the two conditions become one and the same, and we require that $M_{\perp \perp} < 0$ at least some time during inflation. Geometrically, this means that the inflaton must roll along a ridge in the potential for sometime during inflation --- that is, the inflaton trajectory must be unstable.  Perturbations orthogonal to the classical trajectory must result in neighboring trajectories that diverge, giving rise to widely different inflationary dynamics.   Figure 2 illustrates one geometric realization of a potential that meets these criteria.  Moreover, the potential must not only possess a ridge, but the initial conditions must be fine-tuned so that the inflaton rolls along the ridge for a sufficiently long time. Conversely, whenever attractor solutions exist, non-Gaussianity will be small.   
 
Also in the geometric picture, it is necessary that the inflaton trajectory turn somewhat during inflation, as we argued earlier that $|f_{NL}|$ can only be large if the total sourcing of curvature modes by isocurvature modes is non-zero and usually it must also be moderate.  Otherwise, if the sourcing is too weak or too strong, then the trigonometric functions representing the sourcing effects will still force $|f_{NL}|$ to be small.   Interestingly, how much of a turn is needed depends on the relative amplitude of the isocurvature modes during the turn.  If the isocurvature modes are small at the time, then a larger turn is need.  However, if the isocurvature modes are large, then only a minuscule turn in the trajectory is needed to produce moderate sourcing.  

Cumulatively, these results explain why it has been so difficult to realize large non-Gaussianity in two-field inflation.  The potential must be fine-tuned enough to possess a steep ridge, while the initial conditions must be fine-tuned enough that the inflaton rolls along the ridge for a significant length of time during inflation, but also slightly turns.  Nonetheless, a few two-field scenarios that produce large non-Gaussianity have been identified.  Hybrid inflation, where the waterfall field counts as one of the two fields, is a great example of an inflationary scenario possesses a negative entropy mass during inflation and that meets these conditions.  As a second example, Byrnes \textit{et. al.} \cite{ByrnesEtAl-2008} showed that product potentials where one of the two fields dominates the inflationary dynamics (meaning that the kinetic energy of one field is much larger than the kinetic energy of the other) during all of inflation, but the subdominant field picks up speed logarithmically faster than the dominant field also produce large $|f_{NL}|$.   Finally, we note that in investigating sum potentials, many have found it more difficult to find scenarios in which $|f_{NL}|$ is large, but the primary reason for this trouble is that the focus has been on scenarios in which the entropy mass is strictly positive.  Therefore, observable large non-Gaussianity should be more readily achieved in sum potentials by searching for those scenarios that allow the entropy mass to be negative.

\section{Trispectrum}
\label{trispec}

Now we calculate the local trispectrum.  The local trispectrum can be expressed in terms of two dimensionless non-linear parameters, $\tau_{NL}$ and $g_{NL}$:
\begin{align}
T_{\mathcal{R}} & = \tau_{NL} \left[\mathcal{P}_{\mathcal{R}}(|\mathbf{k}_1+\mathbf{k}_3|) \mathcal{P}_{\mathcal{R}}(k_3) \mathcal{P}_{\mathcal{R}}(k_4) + 11 \, \mathrm{perms} \right] \nonumber \\
& + \frac{54}{25} g_{NL} \left[\mathcal{P}_{\mathcal{R}}(k_2) \mathcal{P}_{\mathcal{R}}(k_3) \mathcal{P}_{\mathcal{R}}(k_4) + 3 \, \mathrm{perms} \right]. 
\end{align}
where under the $\delta N$ formalism \cite{AldaiaAndWands-2005,ByrnesEtAl-2006}
\begin{align}
\label{tNL and gNL in terms of delta N}
\tau_{NL} = & \frac{\mathbf{e}_N^T \, \boldsymbol{\nabla} \boldsymbol{\nabla}^T N \, \boldsymbol{\nabla} \boldsymbol{\nabla}^T  N \,  \mathbf{e}_N}{|\boldsymbol{\nabla} N|^4} , \nonumber \\
\frac{54}{25} g_{NL} & = \frac{\mathbf{e}_N^T \mathbf{e}_N^T \mathbf{e}_N^T \, \boldsymbol{\nabla}  \boldsymbol{\nabla}  \boldsymbol{\nabla}  N }{|\boldsymbol{\nabla} N|^3}. 
\end{align}

Since the expression for $g_{NL}$ is less illuminating and cannot be completely expressed in terms of other observables, we focus on $\tau_{NL}$. Equation (\ref{tNL and gNL in terms of delta N}) for $\tau_{NL}$ is equivalent to
\begin{align}
\label{tau as norm squared}
\tau_{NL} = \left|\left| \frac{\boldsymbol{\nabla} \boldsymbol{\nabla}^T N \, \mathbf{e}_N}{|\boldsymbol{\nabla} N|^2} \right|\right|^2.
\end{align}
Using equation (\ref{tau as norm squared}) and the fact that we already calculated the form of the vector $\frac{\boldsymbol{\nabla} \boldsymbol{\nabla}^T N \, \mathbf{e}_N}{|\boldsymbol{\nabla} N|^2}$ in equation (\ref{key delta N step}), we can quickly arrive at the answer.  Dividing equation (\ref{cos Delta para cmpt of key delta N step}) by $\cos \Delta_N$, we obtain the $\boldsymbol{e}_\parallel$ component of $\frac{\boldsymbol{\nabla} \boldsymbol{\nabla}^T N \, \mathbf{e}_N}{|\boldsymbol{\nabla} N|^2}$: 
\begin{align}
\frac{\mathbf{e}_{\parallel} \, \boldsymbol{\nabla} \boldsymbol{\nabla}^T N \, \mathbf{e}_N}{|\boldsymbol{\nabla} N|^2}  = \frac{1}{2} \cos \Delta_N \, (n_{\mathcal{R}}-n_T).
\end{align}  
Similarly, dividing equation (\ref{sin Delta para cmpt of key delta N step}) by $\sin \Delta_N$ gives us the $\mathbf{e}_\perp$ component:
\begin{align}
\frac{\mathbf{e}_\perp \, \boldsymbol{\nabla} \boldsymbol{\nabla}^T N \, \mathbf{e}_N}{|\boldsymbol{\nabla} N|^2}  &  =  \cos \Delta_N  \times \\ 
& \left[M_{\parallel \perp}^* + \sin \Delta_N \cos \Delta_N \sqrt{2\epsilon_*} \, \mathbf{e}_{\perp} \cdot \boldsymbol{\nabla}_* T_{\mathcal{RS}}\right]. \nonumber
\end{align}
Substituting the above two equations into equation (\ref{tau as norm squared}), we find that
\begin{widetext}
\begin{align}
\tau_{NL} = \frac{1}{4} \cos^2 \Delta_N  (n_{\mathcal{R}}-n_T)^2 + \cos^2 \Delta_N \left[M^*_{\parallel \perp} + \sin \Delta_N \cos \Delta_N \sqrt{2\epsilon_*} \, \mathbf{e}_{\perp}^* \cdot \boldsymbol{\nabla} T_{\mathcal{RS}} \right]^2.
\end{align}
\end{widetext}
Using equations (\ref{nR}), (\ref{rC}), and (\ref{fNL general expression}), we can write $\tau_{NL}$ completely in terms of observables, giving the following consistency condition:
\begin{widetext}
\begin{align}
\label{tau in terms of obs}
\boxed{\tau_{NL} = \frac{1}{4} \left(1-r_C^2\right) \, [n_{\mathcal{R}}-n_T]^2 + \frac{1}{r_C^2} \left[\frac{6}{5}f_{NL} + \frac{1}{2} \left(1-r_C^2\right)  (n_{\mathcal{R}}-n_T) \, \right]^2,}
\end{align}
\end{widetext}
where recall that $r_C$ is the curvature-isocurvature correlation.  This gives us a new consistency relation that is unique to two-field inflation and that relates the observables $\tau_{NL}$, $f_{NL}$, $n_{\mathcal{R}}$, $n_T$, and $r_C$. 

Examining equation (\ref{tau in terms of obs}), since the magnitudes of $n_{\mathcal{R}}$ and $n_T$ are constrained to be much smaller than unity, $\tau_{NL}$ can only be large if $\left(\frac{f_{NL}}{r_C}\right)^2$ is large.  Previously, it was shown \cite{SuyamaAndYamaguchi-2007} that
\begin{align}
\label{S-Y bound}
\tau_{NL} \ge \left(\frac{6}{5} f_{NL}\right)^2,
\end{align}
proving that $\tau_{NL}$ will be large whenever $f_{NL}^2$ is large, but this result still left open the question of whether it is possible for $\tau_{NL}$ to be large if $f_{NL}^2$ is not.  This question has since been answered affirmatively for particular models (e.g., \cite{IchikawaEtAl-2008,ByrnesEtAl-2008b}).  Above we show that it is more generally possible for $\tau_{NL}$ to be large even if $f_{NL}^2$ is not, but only if $r_C^2$ is small.  As $r_C = \sin \Delta_N$ reflects the degree of sourcing of curvature modes by isocurvature modes, this shows that the commonality in two-field models where $\tau_{NL}$ is detectably large when $f_{NL}$ is not is that the sourcing effects (the multi-field effects) are weak.

In the limit where $|f_{NL}| \gtrsim 1$, equation (\ref{tau in terms of obs}) reduces to   
\begin{align}
\label{tNLeq}
\tau_{NL} \approx  \frac{1}{r_C^2} \left(\frac{6}{5} f_{NL}\right)^2. 
\end{align}
In this limit, how much $\tau_{NL}$ exceeds the Suyama-Yamaguchi bound in equation (\ref{S-Y bound}) depends only on $r_C$.   While we might naively expect that making $r_C$ as small as possible would maximize the value of $\tau_{NL}$, this is not necessarily the case.  This is because there is a trade-off: scenarios in which the multi-field effects are very small ($r_C \ll 1$) behave in many ways like single-field models and hence they are likely to produce small values for $|f_{NL}|$.  Therefore, reducing $r_C$ even further while preserving larger values for $|f_{NL}|$ typically comes at the expense of even more fine-turning.  Therefore, for detectable non-Gaussianity without excessive fine-tuning, we might expect more typically that $\tau_{NL}$ is no more than one to two orders of magnitude greater than $f_{NL}^2$.

Using a similar approach, an expression for $g_{NL}$ can be found.  However, the result can be expressed only partially in terms of observables.  Below, we cast the result in the most simple and transparent way.  Starting by operating $\frac{1}{|\boldsymbol{\nabla} N|} \, \mathbf{e}_N \cdot \boldsymbol{\nabla}$ on the expression for $f_{NL}$, and then using the definitions of the non-linear parameters and that
\begin{align}
\tau_{NL} = & \frac{\mathbf{e}_N^T \, \boldsymbol{\nabla} (|\boldsymbol{\nabla} N| \mathbf{e}_N^T) \boldsymbol{\nabla} \boldsymbol{\nabla}^T N \, \mathbf{e}_N}{|\boldsymbol{\nabla} N|^4}, \nonumber \\
= & \frac{\mathbf{e}_N^T \, \boldsymbol{\nabla} (|\boldsymbol{\nabla} N|) \, \mathbf{e}_N^T \, \boldsymbol{\nabla} \boldsymbol{\nabla}^T N \, \mathbf{e}_N}{|\boldsymbol{\nabla} N|^4} \\ &  \, \, \, \, \, \, \, \, \, \, \, \, \, \, \, \, + \frac{\mathbf{e}_N^T \, (\boldsymbol{\nabla} \mathbf{e}_N^T) \, \boldsymbol{\nabla} \boldsymbol{\nabla}^T N \, \mathbf{e}_N}{|\boldsymbol{\nabla} N|^3} , \nonumber 
\end{align}
$g_{NL}$ can be written as
\begin{align}
\frac{54}{25}  g_{NL} = - & 2 \tau_{NL} + 4 \left(\frac{6}{5} f_{NL}\right)^2 \nonumber \\ & + \sqrt{\frac{r_T}{8}} \, \mathbf{e}_N \cdot \boldsymbol{\nabla} \left(-\frac{6}{5}f_{NL}\right).
%\frac{54}{25}  g_{NL} = 2 & \left(\frac{6}{5}f_{NL}\right)^2 + 2 \sqrt{\left(\frac{1}{r_c^2} - 1\right)\left(\tau_{NL} - \left(\frac{6}{5} f_{NL}\right)^2\right)} \left(\frac{6}{5} f_{NL}  + \frac{n_{\mathcal{R}} - n_T}{2}\right)  + (1 - r_C^2) \left(-\frac{6}{5}f_{NL}\right)' \nonumber \\ & + r_C \sqrt{(1-r_C^2)(-n_T)} \, \mathbf{e}_{\perp}^* \cdot \boldsymbol{\nabla}_* \left(-\frac{6}{5} f_{NL}\right).
\end{align}
Hence $g_{NL}$ can be large in magnitude only if $\tau_{NL}$ is large, $f_{NL}^2$ is large, and/or if $f_{NL}$ varies dramatically in a small neighborhood about the initial conditions.  Therefore, if $g_{NL}$ is large, but neither $\tau_{NL}$ and $f_{NL}$ are, then it means that the inflaton trajectory is near neighboring trajectories that do produce large $f_{NL}$ and/or that $f_{NL}$ has very strong scale-dependence.

\section{Conclusions}
\label{conclusions}

In this paper, we have derived formulae for the local form of the bispectrum and trispectrum in general two-field inflation.  In particular, we found semi-analytic expressions for $f_{NL}$, $\tau_{NL}$, and $g_{NL}$, the only non-linear parameters in the expressions for the bispectrum and trispectrum whose magnitudes have the potential to be at least of order unity. To do so, we worked within the $\delta N$ formalism, which expresses the bispectrum and trispectrum in terms of gradients of $N$, where $N$ is the number of $e$-folds of inflation.  To perform the calculation, we invoked the slow-roll and slow-turn approximations, and we used a unified kinematical framework, the transfer matrix formalism, and a general expression for the evolution of isocurvature modes. 

We showed that $f_{NL}$ can be written in terms of sines and cosines (related to the degree of sourcing) times $n_{\mathcal{R}} = 1 - n_s$, $n_T$, the turn rate at horizon exit, and $\sqrt{2\epsilon_*} \, \mathbf{e}_{\perp}^* \cdot \boldsymbol{\nabla}_* T_{\mathcal{RS}}$, where $T_{\mathcal{RS}}$ is the transfer function that encodes the relative degree of sourcing of curvature modes by isocurvature modes.  As the magnitudes of all quantities but the term 
%\sin^2(2 \Delta_N)\sqrt{\epsilon_*/8} \, \mathbf{e}_{\perp}^* \cdot \boldsymbol{\nabla} T_{\mathcal{RS}}$ 
$\sin^2 \Delta_N \cos^2 \Delta_N \sqrt{2\epsilon_*} \, \mathbf{e}_{\perp}^* \cdot \boldsymbol{\nabla}_* T_{\mathcal{RS}}$ 
are constrained to be less than unity, $|f_{NL}|$ can only be large when (1) $T_{\mathcal{RS}}$ is extremely sensitive to a change in initial conditions orthogonal to the inflaton trajectory and (2) the total sourcing is non-zero, though usually the total sourcing must also be moderate.  The former condition makes sense on an intuitive level, as to produce a large amount of skew in the primordial perturbations, fluctuations off the classical inflaton trajectory must result in very different inflationary dynamics for the field perturbations.  Now since $T_{\mathcal{RS}}$ is an integral of the turn rate and the relative amplitude of isocurvature modes ($T_{\mathcal{SS}}$), the former condition implies that neighboring trajectories must have dramatically different turn rate profiles, $T_{\mathcal{SS}}$ profiles, or both.  Though we only presented proofs of these conditions for two-field inflation, similar conditions hold for multi-field inflation as well.  

Next, we found an expression for $\sqrt{2\epsilon_*} \mathbf{e}_{\perp}^* \cdot \boldsymbol{\nabla}_* T_{\mathcal{RS}}$ for analytically solvable and similar scenarios that depends on the entropy mass, the isocurvature transfer function $T_{\mathcal{SS}}^e$, and a model-dependent correction $\gamma$, which quantifies the coupling between the fields.  Invoking minimal assumptions about the terms in equation (\ref{grad T_RS final}), we showed that for $\sqrt{2\epsilon_*} \mathbf{e}_{\perp}^* \cdot \boldsymbol{\nabla}_* T_{\mathcal{RS}}$ to be large requires that $M_{\perp \perp} - \gamma$ be large at horizon exit, that the relative amplitude of isocurvature modes ($T_{\mathcal{SS}}$) at the end of inflation be large, and/or that $T_{\mathcal{SS}}^e$ be very sensitive to changes in the initial conditions perpendicular to the inflaton trajectory.

We then further explored the conditions for large non-Gaussianity in general two-field inflation.  After proving an upper bound for $T_{\mathcal{RS}}$ in the case where $T_{\mathcal{SS}} \le 1$ during all of inflation, we argued that if neighboring trajectories do not diverge, then due to constraints on higher-order SRST parameters, the amount of sourcing cannot vary dramatically among neighboring trajectories and hence non-Gaussianity cannot be large.  Therefore, either $T_{\mathcal{SS}} > 1$ or $M_{\perp \perp} < 0$ sometime during inflation, and under some minimal assumptions, these conditions become one and the same.  Geometrically, this means that $f_{NL}$ will be large only if the inflaton traverses along a ridge in the inflationary potential at some point during inflation and the inflaton trajectory at least slightly turns so that the total sourcing of curvature modes by isocurvature modes is moderate.    Unfortunately, though, this implies that some fine-tuning of the potential and/or the initial conditions is needed both to produce a steep enough ridge and/or to situate the inflaton on top of the ridge without it falling off too quickly and yet still slightly turning.   Inflationary scenarios that are attractor solutions therefore cannot produce large $f_{NL}$.    This explains why it has been so difficult to achieve large non-Gaussianity in two-field inflation.  Moreover, it explains why large non-Gaussianity arises in models such as hybrid and multi-brid inflation, axionic $\mathcal{N}$-flation, and tachyonic (p)reheating.  The common denominator of these models is a significant negative curvature (mass) along the entropic direction.  

Finally, we showed that the calculations of $\tau_{NL}$ and $g_{NL}$ are very similar to that of $f_{NL}$.  $\tau_{NL}$ can be written entirely in terms of the spectral observables $f_{NL}$, $n_{\mathcal{R}}$, $n_T$, and $r_C$, where $r_C$ is the dimensionless curvature-isocurvature correlation. This provides a new consistency relation unique to two-field inflation.   Moreover, it sheds new light on the Suyama-Yamaguchi bound $\tau_{NL}  \ge \left(\frac{6}{5} f_{NL}\right)^2$, showing that for $|f_{NL}| \gtrsim 1$, $\tau_{NL} = \left(\frac{6f_{NL}/5}{r_C}\right)^2$.  Though theoretically one could attempt to minimize $r_C$ to make $\tau_{NL}$ even larger relative to $f_{NL}$, this usually comes at the unwanted expense of further fine-turning.  We also calculated the trispectrum parameter $g_{NL}$ and showed that it can only be large in magnitude if $\tau_{NL}$ is large, $f_{NL}^2$ is large, and/or $f_{NL}$ varies dramatically in a small neighborhood about the initial conditions.

Our results for the local bispectrum and trispectrum from inflation allow us to better test and constrain two-field models of inflation using observational data.  Our results also provide better guidance for model-builders seeking to find inflationary models with large non-Gaussianity.  In the future, it will be interesting to explore the range of shapes of ridges that give rise to large non-Gaussianity and to better understand the degree of fine-turning needed in the potential and/or initial conditions.  Finally, it is important to better understand the model-dependent nature of (p)reheating and the aftermath of inflation, to understand the impact on the primordial non-Gaussianity from inflation.

\acknowledgements{The authors wish to thank Christian Byrnes for helpful comments.  This work was supported by an NSF Graduate Research Fellowship, NSF grants AST-0708534 \& AST-0908848, and a fellowship from the David and Lucile Packard Foundation.}

\end{document}